\documentclass[]{spie}  %>>> use for US letter paper
\usepackage{amsmath,amsfonts,amssymb}
\usepackage{graphicx}
\usepackage[colorlinks=true, allcolors=blue]{hyperref}

\usepackage{verbatim}

\title{Towards a response function for the COSI anticoincidence system: preliminary results from Geant4 simulations}

\author[a, b, c]{Alex Ciabattoni}
\author[b]{Valentina Fioretti}
\author[d]{John Tomsick}
\author[d]{Andreas Zoglauer}
\author[e]{Pierre Jean}
\author[e]{Daniel Alvarez Franco}
\author[e]{Peter von Ballmoos}
\author[b]{Andrea Bulgarelli}
\author[a, b]{Cristian Vignali}
\author[b]{Nicolò Parmiggiani}
\author[a, b]{Gabriele Panebianco}
\author[b]{Luca Castaldini}
\affil[a]{Dipartimento di Fisica e Astronomia, Università di Bologna, Via P. Gobetti 93/2, 40129 Bologna, Italy}
\affil[b]{INAF OAS Bologna, Via P. Gobetti 93/2, 40129 Bologna, Italy}
\affil[c]{Fondazione ICSC, Italy}
\affil[d]{Space Sciences Laboratory, University of California, Berkeley, CA 94720, US}
\affil[e]{IRAP, Université de Toulouse, CNRS, CNES, UPS, 9 avenue du colonel Roche, 31028 Toulouse, France}

\authorinfo{Further author information:\\Alex Ciabattoni: E-mail: alex.ciabattoni@inaf.it}

\begin{document} 
\maketitle

\begin{abstract}
The Compton Spectrometer and Imager (COSI) is an upcoming NASA Small Explorer satellite mission scheduled for launch in 2027 and designed to conduct an all-sky survey in the energy range of 0.2-5 MeV. Its instrument consists of an array of germanium detectors surrounded on four sides and underneath by active shields that work as anticoincidence system (ACS) to reduce the contribution of background events in the detectors. These shields are composed of bismuth germanium oxide (BGO), a scintillator material, coupled with Silicon photomultipliers, aimed to collect optical photons produced from interaction of ionizing particles in the BGO and convert them into an electric signal. The reference simulation framework for COSI is MEGAlib, a set of software tools based on the Geant4 toolkit.
The interaction point of the incoming radiation, the design of the ACS modules and the BGO surface treatment change the light collection and the overall shielding accuracy.
The use of the Geant4 optical physics library, with the simulation of the scintillation process, is mandatory for a more realistic evaluation of the ACS performances. However, including the optical processes in MEGAlib would dramatically increase the computing time of the COSI simulations.
We propose the use of a response function encoding the energy resolution and 3D light yield correction based on a separate Geant4 simulation of the ACS that includes the full optical interaction. We present the verification of the Geant4 optical physics library against analytical computations and available laboratory measurements obtained using PMTs as readout device, as a preparatory phase for the simulation of the COSI ACS response.

\end{abstract}


\section{INTRODUCTION}
\label{sec:intro}
The COSI SMEX mission\cite{Tomsick:2023aue}  is a Compton telescope operating in the energy range of 0.2-5 MeV in Low Earth Orbit, with a launch planned in 2027. Its main scientific goals include studies of the 0.511 MeV emission from antimatter annihilation in the Galaxy, mapping radioactive elements from nucleosynthesis, exploiting polarization measurements to determine emission mechanisms and source geometries, and detecting multimessenger sources. Its instrument consists of an array of germanium detectors (GeDs) surrounded on four sides and underneath by scintillation panels that work as anticoincidence system (ACS). These shields produce scintillation light whenever an interaction with ionising radiation occurs. This optical light is then collected and converted into electric signal by readout devices that are coupled to the scintillator crystals. Thanks to this detection capability, the ACS is able not only to passively shield background events, but also to actively raise vetoes for those photons and particles that hit the GeDs and release an energy $>80$ keV in the ACS panels simultaneously. Additionally, thanks to their large field of view, they can work as monitor for gamma transient events, like Gamma-Ray Bursts (GRBs). In fact, a transient event would cause a sudden count rate increase in the ACS data, and when the rate surpasses a predeterminated significance level a trigger is raised. \\ In 2016, the COSI instruments experienced a 46 days long balloon flight\cite{Kierans:2016qik}, allowing the acquisition of the very first scientific data and testing their calibration and general performance. The COSI-2016 instrumentation consisted of 12 GeDs surrounded by cesium iodide (CsI) shield panels on the four sides and underneath. Each ACS panel was coupled to two Photomultipier Tubes (PMTs) for the collection of scintillation light. The COSI SMEX satellite plans to use 16 GeDs and bismuth germanium oxide (BGO) shield panels. The chosen readout devices are now silicon photomultipliers (SiPMs). The SiPMs are generally more compact than PMTs, can be packed into arrays and provide excellent sensibility. \\ The reference framework used for the simulation and data analysis of COSI is the Medium-Energy Gamma-Ray Astronomy library (MEGAlib)\cite{ZOGLAUER2006629}. MEGAlib is an open-source software library designed to simulate and analyze data of gamma-ray telescopes, with a particular focus on Compton telescopes like COSI. It allows for the construction of a virtual detector geometry with \textit{Geomega} and the generation of Monte-Carlo simulated data with \textit{Cosima}. \textit{Revan} and \textit{Mimrec} are instead designed to perform event reconstruction and high-level data analysis, respectively. The COSI simulation pipeline needs to include the real detector effects (e.g. detector resolution, thresholds etc.) that make the simulated data closer to real measurements. The detector effect engine (DEE) is included in the \textit{Nuclearizer} software and applies all these corrections to the \textit{Cosima} simulation. The DEE needs dedicated calibration measurements that would allow the simulation to be properly benchmarked\cite{Zoglauer:2021coa}. \\ The \textit{Cosima} software is based upon Geant4\cite{AGOSTINELLI2003250, 1610988, ALLISON2016186}. Although developed for X-ray and Gamma-ray simulations, Geant4 also allows the simulation of optical scintillation, i.e. when a material (the scintillator) emits light in the visible and ultraviolet band under excitation from ionising radiation. When a cosmic-ray particle or a photon deposits energy in the ACS module, the
scintillation photons (thousands for a deposit of hundreds of keV) scatter within the material and eventually some of them reach the readout device and are converted into an electric pulse that is then converted back into energy with a calibrated energy-pulse
height relation. In \textit{Cosima}, the generation and transport of the optical photons are not simulated and the energy
deposit by the high-energy photon or particle is directly recorded. While including the production and tracking of the optical photons would increase the accuracy of the simulated efficiency and energy resolution, it would also dramatically increase the computational time. To account for this, we propose to use dedicated Geant4 simulations, including the optical process, to model the spatial end energy distribution of the light collection into a response function, to be included in the COSI simulation DEE. In this perspective, we present here a preliminary verification of the Geant4 optical physics, using both analytical models and laboratory measurements to be compared with the simulation results. The analytical model is based on Snell and Fresnel equations regulating the transmission and reflection probabilities for an optical photon in a two-media surface, while laboratory measurements of the light response of a CsI panel coupled with a PMT are used to test the Geant4 optical simulation in a more complex and realistic scenario. As simulation framework, we use BoGEMMS-HPC\cite{2012SPIE.8453E..35B}, an astronomy-oriented Geant4-based application. \\ This proceedings paper is organized as follows: Sec. \ref{sec:opt_phys} describes the models implemented in the Geant4 optical physics library; in Sec. \ref{sec:an_ver} we show the verification of the Geant4 optical physics against the analytical model; in Sec. \ref{sec:claire} we present the comparison of Geant4 simulation with the laboratory measurements; finally, Sec. \ref{sec:towards} shows preliminary results for the characterization of the response and energy resolution of the COSI SMEX ACS, in view of the future construction of an accurate response function to be implemented in the detector effect engine.

\section{GEANT4 OPTICAL PHYSICS}
\label{sec:opt_phys}
The Geant4 optical physics library\cite{Geant4manual} enables the generation of radiation in the optical band through scintillation. The transport of optical photons through matter is mainly based on two processes: absorption and medium boundary effects. The former is parametrized by the \textit{absorption length} of the scintillator, representing the mean path of an optical photon in that medium. The boundary effects are regulated by a set of parameters that characterize the optical property of the surface and the reflection and refraction probabilities for the optical photons. All these parameters are defined in what is called \textit{Unified Model}, which applies to dielectric-dielectric interfaces and represents the one used in our simulation. The scintillator material is coated with a reflective material and optical photons are either reflected or absorbed. This type of interface is called \textit{painted surface}. There are two main characteristics which discriminate different types of painted surfaces: the finish (\textit{polished} or \textit{ground}) and the presence or absence of a gap between volume and paint (\textit{frontpainted} and \textit{backpainted}). Hence, one can have four different combinations: \textit{polishfrontpainted}, \textit{groundfrontpainted}, \textit{polishbackpainted} and \textit{groundbackpainted}. The level of roughness is parametrized by the parameter $\sigma_\alpha$. A rough surface is modelled as consisting of micro facets, each of which has an angle ($\alpha$) between its normal and the average surface normal sampled from a Gaussian distribution, with mean given by the average surface normal and standard deviation of $\sigma_\alpha$. Hence, a surface with higher $\sigma_\alpha$ is a rougher surface. Another key property is the reflective type. There are four types of reflection that an optical photon can undergo: specular spike (SS), specular lobe (SL), backscattering (BS) and Lambertian (L) reflection. In the verification work against laboratory measurements, we will test SS reflection (i.e. the incidence angle is equal to the reflection angle), and L reflection (i.e. the reflection angle follows a Lambert cosine distribution). The main properties of the four types of surfaces are summarized as follows:
\begin{itemize}
    \item \textbf{PolishedFrontPainted (PFP)} The optical photon can only be reflected or absorbed. The probability of being reflected is set by the parameter Reflectivity; if reflection occurs, then it is specular spike reflection.
    \item \textbf{GroundFrontPainted (GFP)} The optical photon can only be reflected or absorbed. The probability of being reflected is set by the parameter Reflectivity; if reflection occurs, then it is Lambertian reflection.
    \item \textbf{PolishedBackPainted (PBP)} There is an air gap between the crystal and the paint, the latter being a perfectly smooth mirror. The user must set $\sigma_\alpha$, which is associated to the crystal-gap interface, and a refractive index, which is associated to the gap and then used to apply the Snell’s law between the crystal and the gap. If reflection occurs, it can be SS, SL, BS or L according to their assigned probabilities. If refraction occurs, the optical photon enters the gap and interacts with the paint. It can then be either SS reflected or absorbed according to the parameter Reflectivity as in the previous cases. If reflected, the photon interacts again with the gap-crystal interface and Snell’s law is again applied. At this point, the photon can be refracted into the crystal, or reflected back and interact again with the paint.
    \item \textbf{GroundBackPainted (GBP)} As for PBP, there is an air gap between the crystal and the paint, the latter being now a ground mirror instead of a polished one. The user must set $\sigma_\alpha$, which is associated to the crystal-gap interface, and a refractive index, which is associated to the gap and then used to apply the Snell’s law between the crystal and the gap. The reflection can be SS, SL, BS or L according to their assigned probabilities. If refraction occurs, the optical photon enters the gap and can then be either L reflected or absorbed from the paint according to the parameter Reflectivity. If reflected, the photon interacts again with the gap-crystal interface and Snell’s law is again applied. In summary, it is the same as PBP, but the finish of the mirror is ground and the reflection type is Lambertian.

\end{itemize}

%\textbf{COSI, ballon e SMEX, ACS, BGO, PMT e SiPM, MEGAlib e Geant4 optical physics}

\section{VERIFICATION WITH ANALYTICAL MODELS}
\label{sec:an_ver}
We built a toy model for the COSI ACS consisting of a set of layers approximating the interface between a BGO-based scintillator and a PMT, and studied the transmission, reflection and absorption probabilities as functions of the angle of incidence.

\subsection{Geant4 simulation setup}
\label{sec:sim}
\begin{figure}
   \begin{center}
   \begin{tabular}{c} 
   \includegraphics[height=3.5cm]{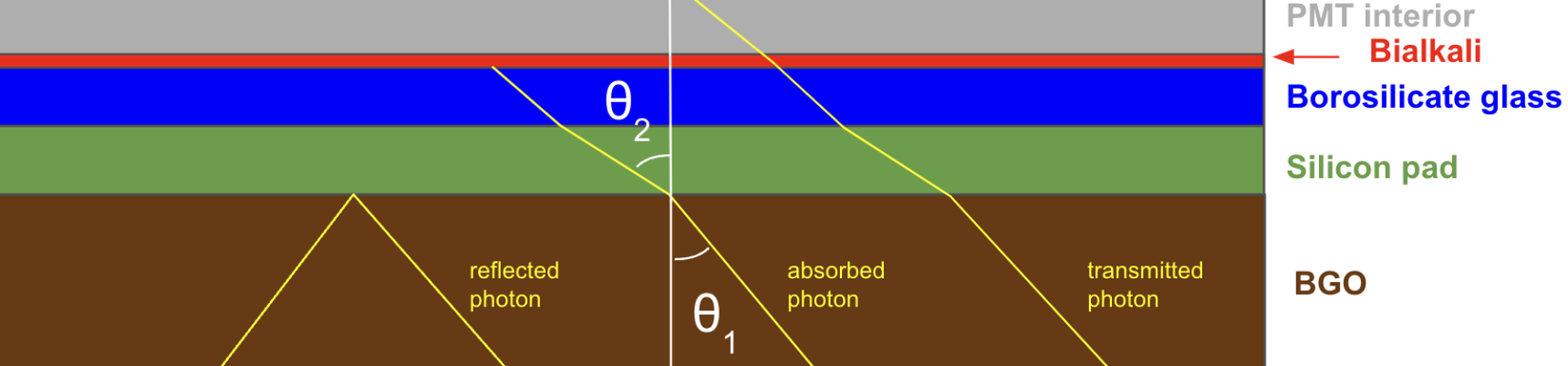}
   \end{tabular}
   \end{center}
   \caption 
   { \label{fig:scheme} 
Schematic view of the propagation of optical photons through the COSI ACS toy model.}
   \end{figure}

The mass model, shown schematically in Fig. \ref{fig:scheme}, consists of a five-layered slab composed as follows:
\begin{enumerate}
    \item the first layer is the BGO, which is an inorganic scintillator made of bismuth germanium oxide (Bi$_4$Ge$_3$O$_{12}$); it has a thickness of 2 cm;
    \item the second layer used as optical coupler between BGO and PMT is the silicon pad, polymer made of siloxane (R$_2$SiOSiR$_2$) and with a thickness of 5 mm;
    \item the third layer is the borosilicate glass, made of SiO2 (80 \%), B$_2$O$_3$ (13\%), Na$_2$O (4\%) and Al$_2$O$_3$ (3\%); it is used as a window on the PMT and it has a thickness of 5 mm;
    \item the fourth layer is the bialkali photocathode, which is made of potassium, cesium and antimony (KCsSb) and it is where a fraction of the optical photons is absorbed and generates photoelectrons; it has a thickness of 20 nm;
    \item the fifth layer is the PMT interior, which is just a slab with vacuum inside; its thickness is set to 1 cm.
\end{enumerate}
The lateral dimension of the slab was set to be quite large (1 m) to prevent possible boundary effects in the simulation. The simulated optical properties of all the materials were specified by their (complex) refractive index, which is listed in Table \ref{tab:rindex} as a function of the wavelength. 
\begin{table}[ht]
\begin{center}       
\begin{tabular}{|c|c|c|c|c|c|c|}
\hline
\rule[-1ex]{0pt}{3.5ex}  \textbf{Wavelength (mm)} & \textbf{BGO} & \textbf{SiPad} & \textbf{Glass} & \textbf{Bialkali} & \textbf{PMT}  \\
\hline
\rule[-1ex]{0pt}{3.5ex}  380 & 2.244 & 1.42 & 1.514 & $1.92 + 1.69\,i$ & 1   \\
\hline
\rule[-1ex]{0pt}{3.5ex}  395 & 2.224 & 1.42 & 1.514 & $2.18 + 1.69\,i$ & 1   \\
\hline
\rule[-1ex]{0pt}{3.5ex}  410 & 2.208 & 1.42 & 1.515 & $3.38 + 1.71\,i$ & 1   \\
\hline
\rule[-1ex]{0pt}{3.5ex}  425 & 2.193 & 1.42 & 1.516 & $2.61 + 1.53\,i$ & 1   \\
\hline
\rule[-1ex]{0pt}{3.5ex}  440 & 2.181 & 1.42 & 1.516 & $2.70 + 1.50\,i$ & 1   \\
\hline
\rule[-1ex]{0pt}{3.5ex}  455 & 2.169 & 1.42 & 1.517 & $2.87 + 1.44\,i$ & 1   \\
\hline
\rule[-1ex]{0pt}{3.5ex}  470 & 2.159 & 1.42 & 1.517 & $3.00 + 1.34\,i$ & 1   \\
\hline
\rule[-1ex]{0pt}{3.5ex}  485 & 2.151 & 1.42 & 1.518 & $3.00 + 1.11\,i$ & 1   \\
\hline
\rule[-1ex]{0pt}{3.5ex}  500 & 2.143 & 1.42 & 1.519 & $3.00 + 1.06\,i$ & 1   \\
\hline
\rule[-1ex]{0pt}{3.5ex}  515 & 2.135 & 1.42 & 1.520 & $3.09 + 1.05\,i$ & 1   \\
\hline 
\rule[-1ex]{0pt}{3.5ex}  530 & 2.129 & 1.42 & 1.520 & $3.26 + 0.86\,i$ & 1   \\
\hline
\rule[-1ex]{0pt}{3.5ex}  545 & 2.123 & 1.42 & 1.521 & $3.20 + 0.63\,i$ & 1   \\
\hline
\rule[-1ex]{0pt}{3.5ex}  560 & 2.118 & 1.42 & 1.522 & $3.12 + 0.53\,i$ & 1   \\
\hline
\rule[-1ex]{0pt}{3.5ex}  575 & 2.113 & 1.42 & 1.524 & $3.06 + 0.46\,i$ & 1   \\
\hline
\rule[-1ex]{0pt}{3.5ex}  590 & 2.110 & 1.42 & 1.525 & $3.01 + 0.42\,i$ & 1   \\
\hline
\rule[-1ex]{0pt}{3.5ex}  605 & 2.104 & 1.42 & 1.526 & $2.98 + 0.38\,i$ & 1   \\
\hline
\rule[-1ex]{0pt}{3.5ex}  620 & 2.100 & 1.42 & 1.528 & $2.96 + 0.37\,i$ & 1   \\
\hline
\rule[-1ex]{0pt}{3.5ex}  635 & 2.097 & 1.42 & 1.530 & $2.95 + 0.35\,i$ & 1   \\
\hline
\rule[-1ex]{0pt}{3.5ex}  650 & 2.094 & 1.42 & 1.532 & $2.95 + 0.34\,i$ & 1   \\
\hline
\rule[-1ex]{0pt}{3.5ex}  665 & 2.091 & 1.42 & 1.534 & $2.95 + 0.34\,i$ & 1   \\
\hline
\rule[-1ex]{0pt}{3.5ex}  680 & 2.088 & 1.42 & 1.515 & $2.96 + 0.33\,i$ & 1   \\
\hline
\end{tabular}
\end{center}
\caption{Refractive indices of all the materials for different photon wavelengths\cite{Alvarez:2022}.} 
\label{tab:rindex}
\end{table} 
As can be noted, only the bialkali photocathode has a non-vanishing imaginary refractive index, which means that it is the only absorbing material in the simulation. In the Geant4 simulation, this absorbing property was taken into account not through the use of a complex refractive index, but rather by setting an (energy-dependent) absorption length ($d$) to the bialkali material, which is related to the imaginary refractive index ($k$) in the following way:
\begin{equation}
    k = \frac{\lambda}{4\pi\,d}
\end{equation}
where $\lambda$ is the photon wavelength in vacuum.
Physically, the absorption length represents the mean free path for the photon before being absorbed in the medium. \\
The model used for all the optical surfaces, which were all treated as polished, was the unified model.

\subsection{The analytical model}
\label{sec:an}
The fundamental equation which relates the incident angle $\theta_1$ and the refractive angle $\theta_2$ (with respect to the vertical line perpendicular to the incidence surface) of the light passing through two different media is the Snell’s law:
\begin{equation}
    \sin{\theta_1}\cdot n_1^* = \sin{\theta_2}\cdot n_2^*
    \label{eq:Snell}
\end{equation}
where $n_1^*$ and $n_2^*$ are the complex refractive indices of the two materials. From Eq. \eqref{eq:Snell} it is possible to express the cosine of $\theta_2$ as follows:
\begin{equation}
    \cos{\theta_2} = \sqrt{1 - \left(\frac{n_1^*}{n_2^*}\cdot\sin{\theta_1}\right)^2}\,\,.
\end{equation}
We note that when $\sin{\theta_1} > (n_2^*/n_1^*)$ the cosine of $\theta_2$ is purely imaginary, i.e. no refraction is possible and total internal reflection takes place. The smallest incidence angle for which we have total internal reflection is called \textit{critical angle} ($\theta_\text{crit}$), that means that for $\theta_1 > \theta_\text{crit}$ reflection always occurs. \\
The transmission and reflection probabilities are calculated through the amplitude coefficients $r$ and $t$, which are polarization-dependent and specified by the following Fresnel’s equations:
\begin{eqnarray}
    r_s &=& \frac{n_1^*\cdot\cos{\theta_1} - n_2^*\cdot\cos{\theta_2}}{n_1^*\cdot\cos{\theta_1} + n_2^*\cdot\cos{\theta_2}} \nonumber\\
    r_p &=& \frac{n_2^*\cdot\cos{\theta_1} - n_1^*\cdot\cos{\theta_2}}{n_1^*\cdot\cos{\theta_2} + n_2^*\cdot\cos{\theta_1}} \nonumber\\
    t_s &=& 1 + r_s \nonumber\\
    t_p &=& \frac{n_1^*}{n_2^*}\cdot (1 + r_p)
    \label{eq:Fresnel}
\end{eqnarray}
where the subscripts $p$ and $s$ refer to parallel and perpendicular polarization with respect to the plane of incidence, respectively. \\
In a system of three materials, photons which are generated in the material 1 and are transmitted to the material 2 can undergo multiple internal reflections before being transmitted to the material 3. Then, the amplitude coefficients for such three-layered configuration ($T$ and $R$) are
\begin{eqnarray}
    T_{(s,p)}(\lambda, \theta_1) &=& \frac{t_{(s,p)12}\cdot t_{(s,p)23}\cdot e^{i\delta_2\cos^2{\theta_2}}}{1 + r_{(s,p)12}\cdot r_{(s,p)23}\cdot e^{i(2\delta_2-\delta_3)}}\,\,, \nonumber \\
    R_{(s,p)}(\lambda, \theta_1) &=& r_{(s,p)12} + \frac{t_{(s,p)12}\cdot t_{(s,p)21}\cdot r_{(s,p)23}\cdot e^{i(2\delta_2-\delta_1)}}{1 + r_{(s,p)12}\cdot r_{(s,p)23}\cdot e^{i(2\delta_2-\delta_1)}}\,\,,
    \label{eq:t_r_amp}
\end{eqnarray}
where
\begin{eqnarray}
    \delta_1 &=& 4\pi\frac{h_2}{\lambda}n_1^*\frac{\sqrt{1-\cos^2{\theta_1}}\cdot\sqrt{1-\cos^2{\theta_2}}}{\cos{\theta_2}}\,\,,\nonumber\\
    \delta_2 &=& \frac{2\pi h_2 n_2^*}{\lambda\cos{\theta_2}}\,\,,\nonumber\\
    \delta_3 &=& 4\pi\frac{h_2}{\lambda}n_3^*\frac{\sqrt{1-\cos^2{\theta_3}}\cdot\sqrt{1-\cos^2{\theta_2}}}{\cos{\theta_2}}\,\,,
\end{eqnarray}
are the phase differences; $h_2$ refers to the thickness of the intermediate layer and $\lambda$ is the photon wavelength (both of them must have the same measurement unit). Such phases take into account the wave nature of light and the fact that it can take paths with different lengths before reaching the material 3, depending on how many reflections occur inside the material 2; this translates in differences to the phases and hence interference effects. \\
Finally, the total reflection and transmission probability for a three-layered slab, equally weighting on both polarizations, are
\begin{eqnarray}
    T(\lambda, \theta_1) &=& 0.5\cdot \left(T_s\cdot \bar{T}_s + T_p\cdot \bar{T}_p\right)\cdot\frac{n_3\cdot\mathbb{R}(\cos{\theta_3})}{n_1\cdot\mathbb{R}(\cos{\theta_1})}\,\,,\nonumber\\
    R(\lambda, \theta_1) &=& 0.5\cdot \left(R_s\cdot \bar{R}_s + R_p\cdot \bar{R}_p\right)\,\,.
\end{eqnarray}
where $\mathbb{R}(x)$ is the real part of $x$. For a system of five materials, it is possible to consider it as composed of two three-layered blocks, with the material 3 of the first one being also the material 1 for the second one (so for our case, BGO-Silicon pad-glass is the first block, glass-bialkali-PMT is the second one). In this perspective, the total probability for a photon to reach the final layer is that it is both transmitted in the first and second block, while the probability of being reflected is that it is reflected in the first block, or it is transmitted in the first block and then reflected in the second one. Hence
\begin{eqnarray}
    T(\lambda, \theta_1) &=& T_1(\lambda, \theta_1)\cdot T_2(\lambda, \theta_1^\prime) \,\,,\nonumber \\
    R(\lambda, \theta_1) &=& R_1(\lambda, \theta_1) + T_1(\lambda, \theta_1)\cdot R_2(\lambda, \theta_1^\prime)\,\,.
\end{eqnarray}
where the incident angle of the second block ($\theta_1^\prime$) is actually the transmission angle from the first one. Since all probabilities must sum up to 1, the absorption is then just
\begin{equation}
    A(\lambda, \theta_1) = 1 - T(\lambda, \theta_1) - R(\lambda, \theta_1)\,\,.
\end{equation}
Once a photon reaches the photocathode, it might be absorbed and generate a photoelectron. The probability that such an event occurs is then
\begin{equation}
    P_{phe}(\lambda, \theta_1) = T(\lambda, 0^\circ)\cdot \epsilon_Q(\lambda, 0^\circ)\cdot \frac{A(\lambda, \theta_1)}{A(\lambda, 0^\circ)}\,\,.
\end{equation}
The first term takes into account the probability that the photon is transmitted from the first block and hence actually reaches the photocathode; the second term is the quantum efficiency\cite{Hamamatsu} at $\theta_1 = 0^\circ$ for the bialkali photocathode; the third term (the ratio of the two absorption terms) encodes the angular dependence which is indeed absent in the quantum efficiency.

\subsection{Results}
\label{sec:res_analyt}
\begin{figure}
   \begin{center}
   \begin{tabular}{cc} 
   \includegraphics[height=6cm]{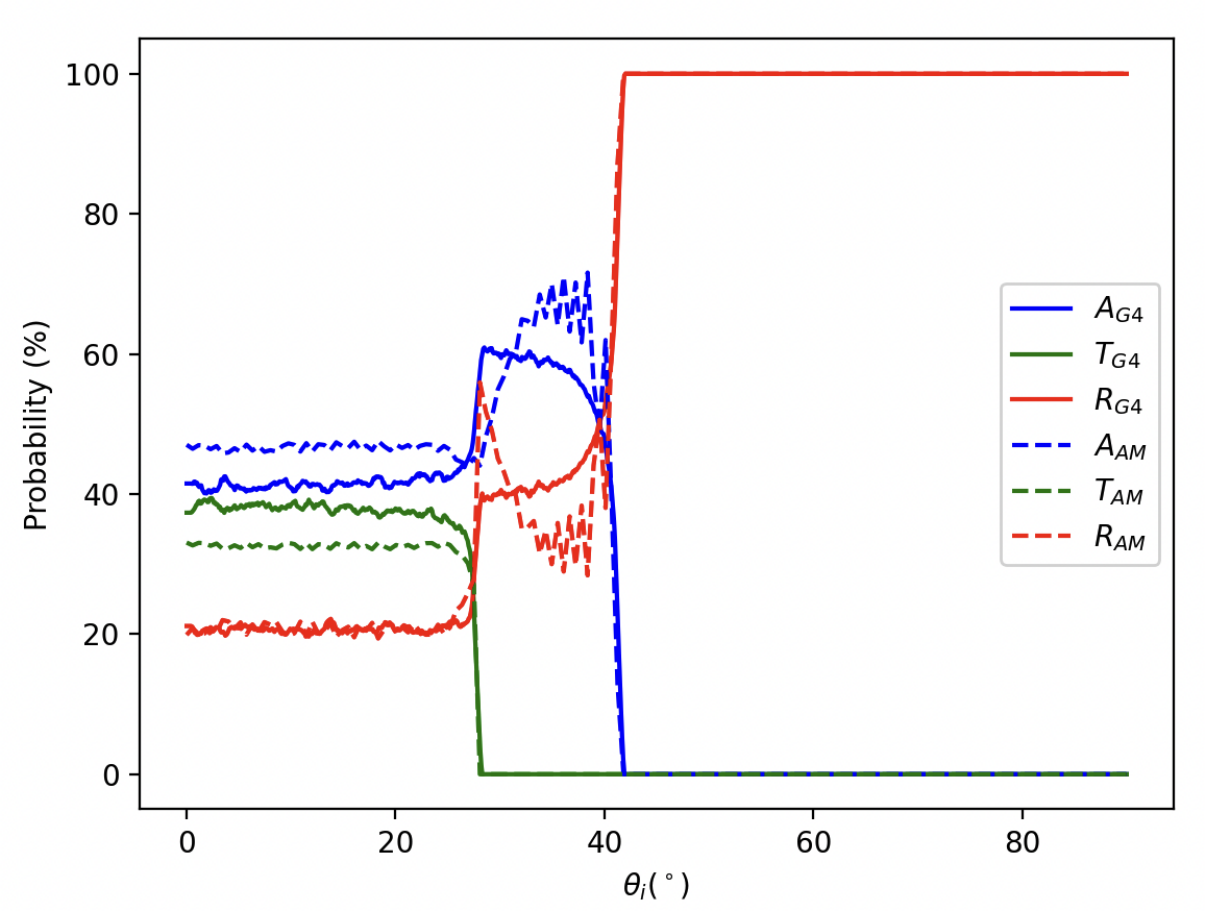}\includegraphics[height=6cm]{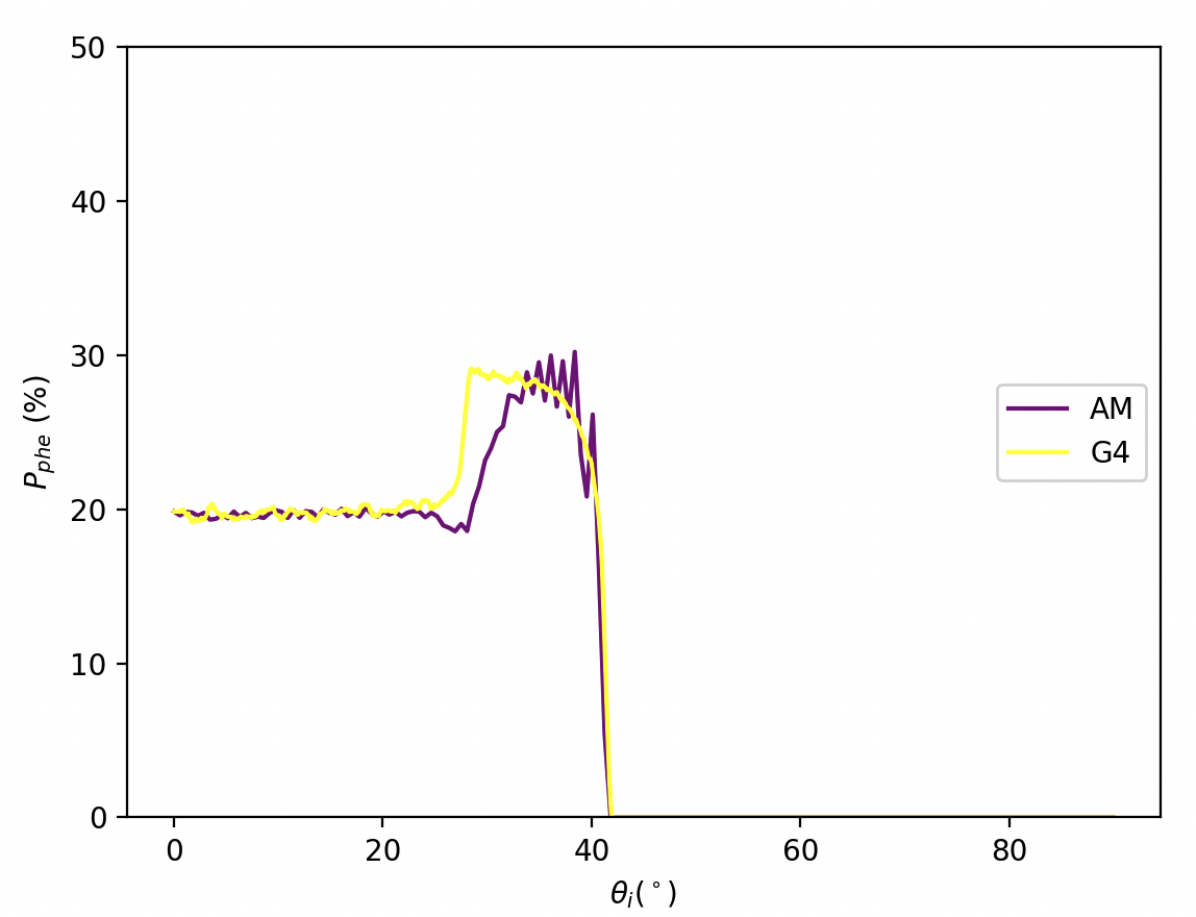}
   \end{tabular}
   \end{center}
   \caption 
   { \label{fig:RTA} 
On left panel, transmission, reflection and absorption probabilities as functions of the incidence angle for the PMT toy model from both the simulation (G4) and the analytical model (AM) (left). On right panel, probability for the emission of a photoelectron as a function of the incidence angle, from both the simulation (G4) and the analytical model (AM).}
   \end{figure}

\begin{figure}
   \begin{center}
   \begin{tabular}{cc} 
   \includegraphics[height=6cm]{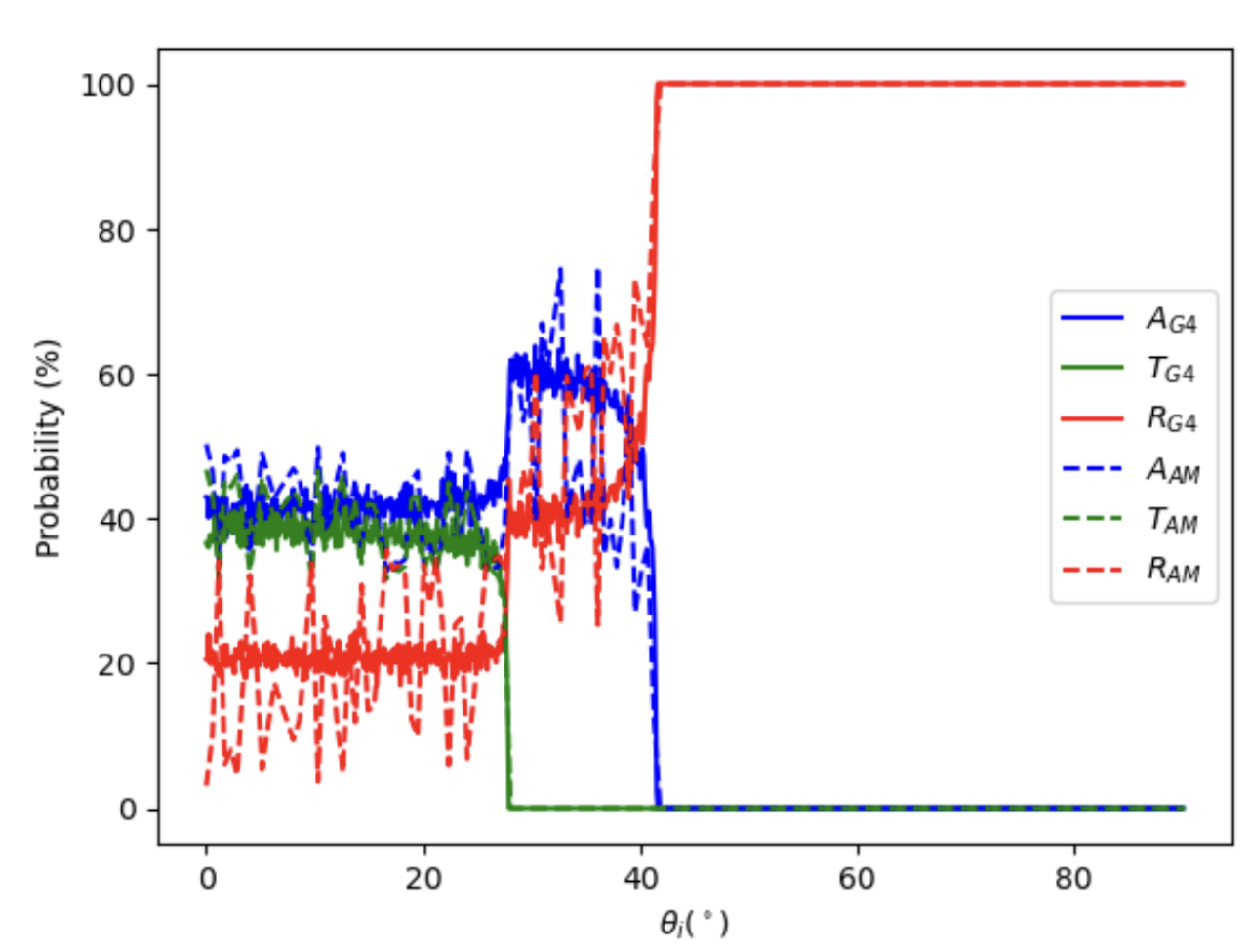}\includegraphics[height=6cm]{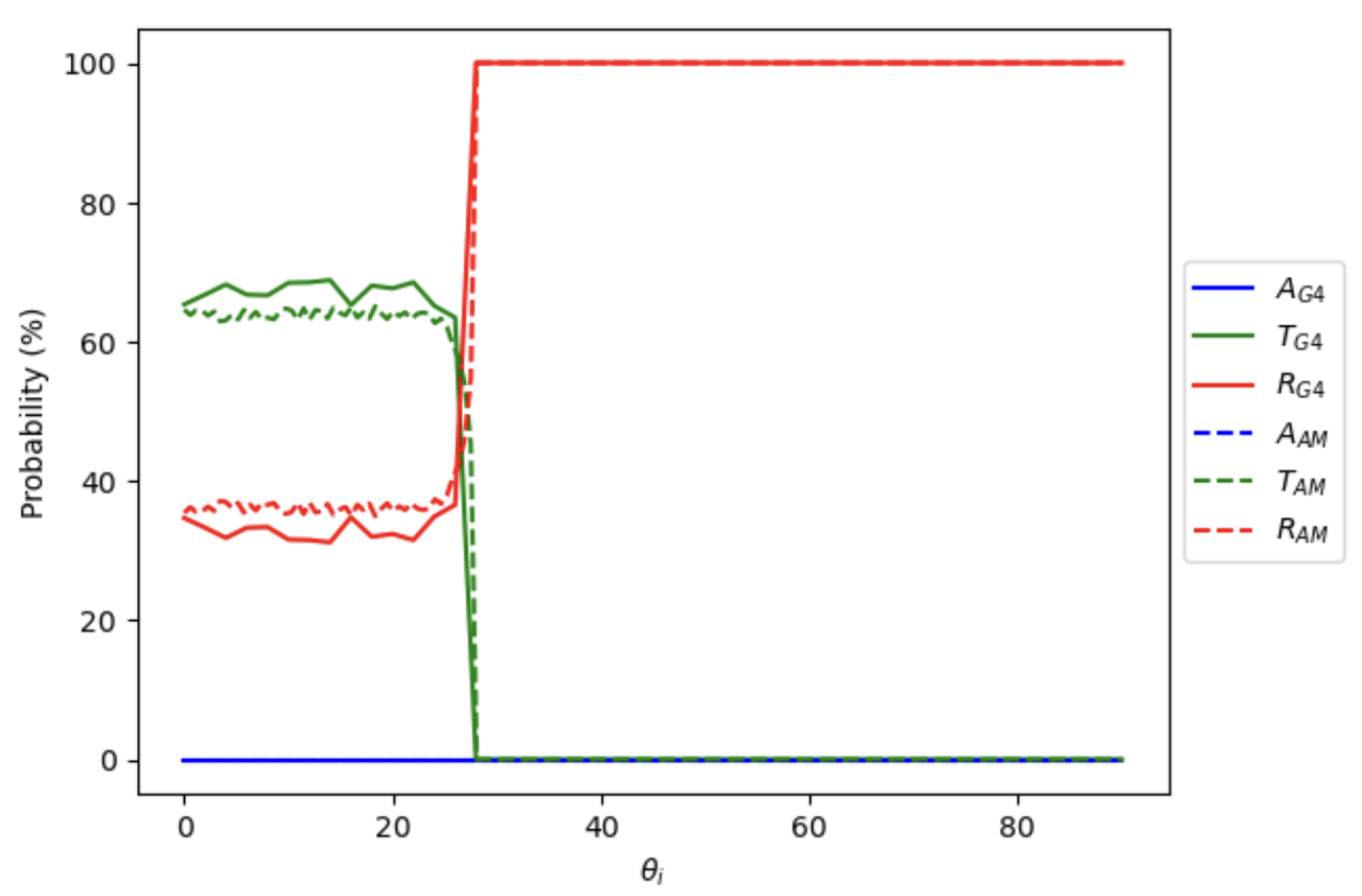}
   \end{tabular}
   \end{center}
   \caption 
   { \label{fig:RTA_mod} 
On the left, transmission, reflection and absorption for the COSI ACS model as functions of the incidence angle for a bialkali thickness  and absorption length both 100 times larger than the standard ones; on the right, the same quantities  with no absorption in the photocathode.}
   \end{figure}

For each simulation, we generated in the BGO 1000 optical photons at 500 nm (2.48 eV) with a fixed angle of incidence with respect to the BGO-SiPad interface, varying the latter from $0^\circ$ to $90^\circ$ at a step of $0.1^\circ$ and obtaining each time the number of transmitted, reflected and absorbed photons. The transmitted (T) photons are those that reach the PMT interior, the reflected (R) ones are the photons that are reflected back into the BGO layer. The absorbed (A) photons are the ones absorbed in the photocathode and the only that can generate photoelectrons (and hence a signal in the PMT). For both analytical and simulated probabilities, we have $T+R+A=1$. \\
The left panel of Fig. \ref{fig:RTA} shows the transmission, reflection and absorption probabilities obtained by the simulation and the analytical model as a function of the incidence angle, while the probability of emission of photoelectrons is instead plotted in the right panel Fig. \ref{fig:RTA}. The Geant4 simulation reproduces the behaviour of the theoretically predicted probabilities, presenting however some deviation from the analytical model. As can be noticed, the transmission probability drops at an incidence angle of $\sim27.3^\circ$. In fact, when $\theta_1$ is larger than such angle, the optical photons will always arrive at the bialkali-PMT boundary with an angle greater than the critical value of the surface and hence they are reflected. The absorption probability is nearly constant at $\sim40\%$ for $\theta_1 \lesssim 27.3^\circ$, then increases up to $\sim60\%$ and finally drops to 0\% at $\sim41^\circ$, that corresponds to the critical angle of the BGO-SiPM interface. For $\theta_1$ larger than this angle, only reflection takes place.
The deviation between Geant4 and the analytical prediction is probably caused by having a very thin layer (the bialkali photocathode) that has a thickness of the same order of the photon wavelength. In fact, in this condition the wave nature of optical photons becomes relevant, and interference effects and frustrated transmission (i.e. transmission through a thin layer) may not be completely captured by the Geant4 simulation, leading to the observed discrepancy. Further investigation of this issue was made by reproducing the same simulation at first with a larger thickness of the photocathode (20 mm) and a proportionally larger absorption length (otherwise the optical photons would always be absorbed in such thick layer). Secondly we removed the absorbing property of the photocathode, that is, without assigning to it an absorption length (keeping its thickness at 20 nm). The results are shown in Fig. \ref{fig:RTA_mod}, where it can be seen that, in the former case, the simulation seems to correctly represent the “average” behaviour of the analytical curve, whereas in the latter case no relevant deviations are observed. The fluctuations in the analytical model are due to the large interference phases in Eq. \eqref{eq:t_r_amp}, since we increased the parameter $h_2$. An important remark is that Geant4 is meant to deal mostly with high energy photons, whose wave nature can be safely neglected, but in situations where optical photons interact with very thin layers the particle description of light may not be a satisfactory approximation. However, despite the lack of such physics in the simulation, Geant4 is observed to not dramatically deviate from the analytical model, and the transmission, reflection and absorption probabilities are still quite close to their theoretically expected values, at every incidence angle.
In Table \ref{tab:RTA} we list the three probabilities after integrating them over the incidence angle, for both the simulation and the analytical model. Geant4 is capable of reproducing the analytical model with an overall accuracy of $<15\%$. In particular, the discrepancy for the number of absorbed photons, particularly important for predicting the number of photoelectrons emitted from the photocathode, is $\sim8\%$.
\begin{table}[ht]
\begin{center}       
\begin{tabular}{|c|c|c|}
\hline
\rule[-1ex]{0pt}{3.5ex}  & \textbf{Analytical model} & \textbf{Geant4 simulation} \\
\hline
\rule[-1ex]{0pt}{3.5ex}  \textbf{Transmission} & 10.0\% & 11.6\% \\
\hline
\rule[-1ex]{0pt}{3.5ex}  \textbf{Reflection} & 66.7\% & 67.1\% \\
\hline
\rule[-1ex]{0pt}{3.5ex}  \textbf{Absorption} & 23.2\% & 21.4\% \\
\hline
\end{tabular}
\end{center}
\caption{ Integrated transmission, reflection and absorption probabilities for both the analytical model and the Geant4 simulation.} 
\label{tab:RTA}
\end{table} 

\section{VERIFICATION AGAINST LABORATORY MEASUREMENTS: THE CLAIRE EXPERIMENT}
\label{sec:claire}
We simulated the interaction of X-ray photons with a block of a scintillator material (CsI(Tl)) coupled with a PMT to estimate the PMT relative response for different beam positions and then compared it with the experimental measurements described in \citenum{Alvarez:2022}.

\subsection{Experimental setup}
\label{sec:expsetup}
\begin{figure}
   \begin{center}
   \begin{tabular}{c} 
   \includegraphics[height=6cm]{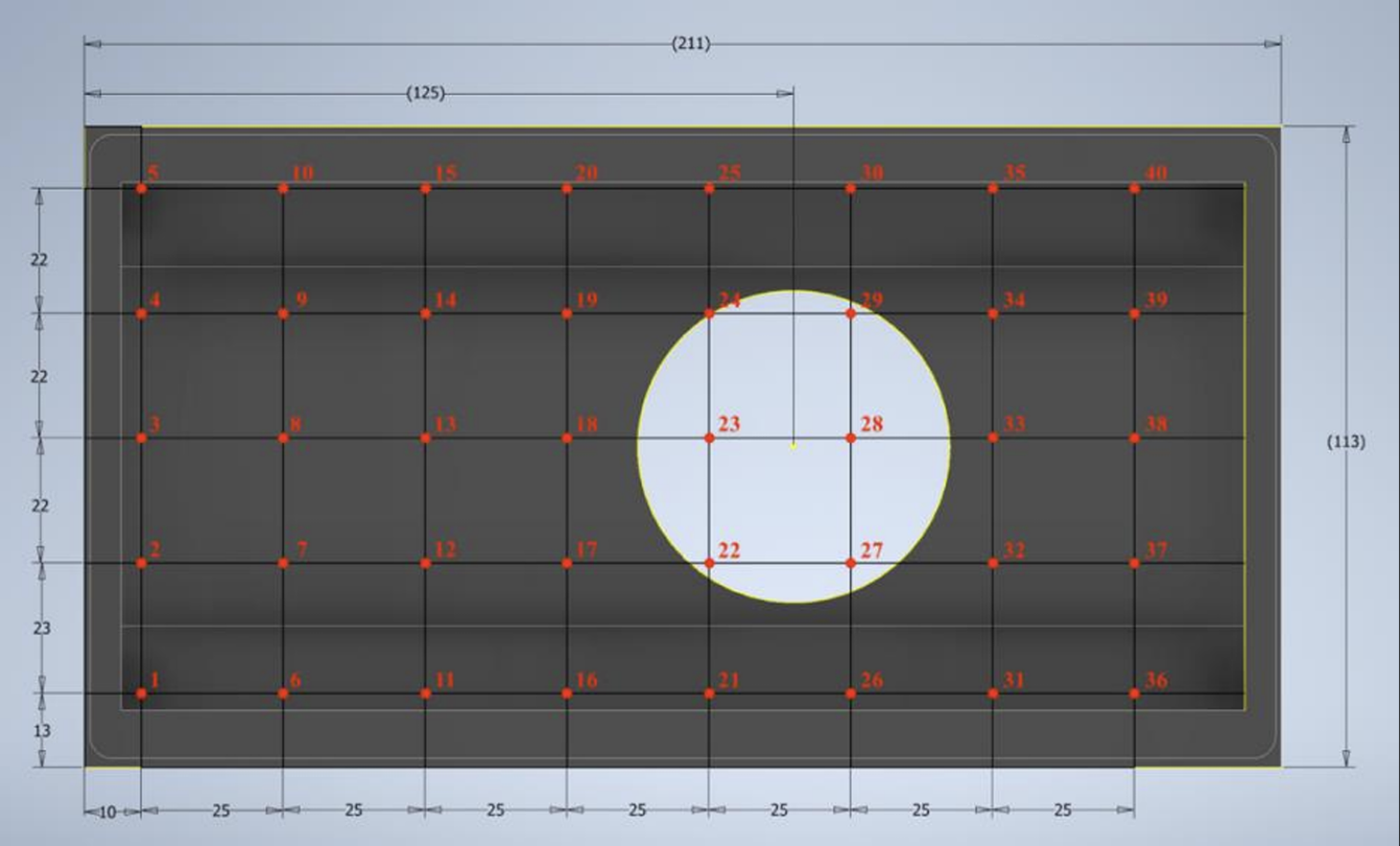}
   \end{tabular}
   \end{center}
   \caption 
   { \label{fig:positions} 
Interaction positions of the 122 keV photons in the bottom face of the CsI crystal and corresponding labels\cite{Alvarez:2022}. The white circle corresponds to the PMT window.}
   \end{figure}

The experiment setup consists of a trapezoidal block of CsI(Tl), with a length of 198 mm, height of 20 mm, and width of 93 mm (for the larger side on the top) and 70 mm (for the smaller side on the bottom), placed inside an aluminum case. The block of CsI is wrapped with 2 layers of Teflon and 2 layers of aluminum foil, whose role is to reflect and keep scintillation photons inside the scintillator. An open window with a diameter of 55 mm in the aluminum case is designed to allow the optical coupling of a PMT with the CsI block, through a silicon pad. Finally, a collimated radioactive source ($^{57}$Co) was placed in front of the bottom side of the aluminum case and 40 measurements of the 122 keV decay line were taken. Each time, the radiation hits the block in a different position, whose distribution and indices are shown in Fig. \ref{fig:positions}, and optical scintillation photons are consequently produced inside the CsI crystal; finally, the PMT responses (proportional to the number of scintillation photons that reach the PMT) were recorded for each position.

\subsection{Geant4 simulation setup}
\label{sec:simsetup}
Concerning the simulated mass model, the geometries of the CsI block and the aluminum case were imported using the CADMesh library\cite{poole2012acad}, whereas all the other components were constructed as Geant4 standard volumes. The crystal-PMT interface was simulated as follows: a silicon pad (SiPad) (with a thickness of 5 mm) optically couples the crystal to the borosilicate glass (with the same thickness) within the open window; above the glass, there is the photocathode (with a thickness of 20 nm) and finally the PMT (with a height of 2 mm). All the layers are cylinders with a diameter of 46 mm.

\subsection{Methods}
\label{sec:methods}
\begin{figure}
   \begin{center}
   \begin{tabular}{c} 
   \includegraphics[height=6cm]{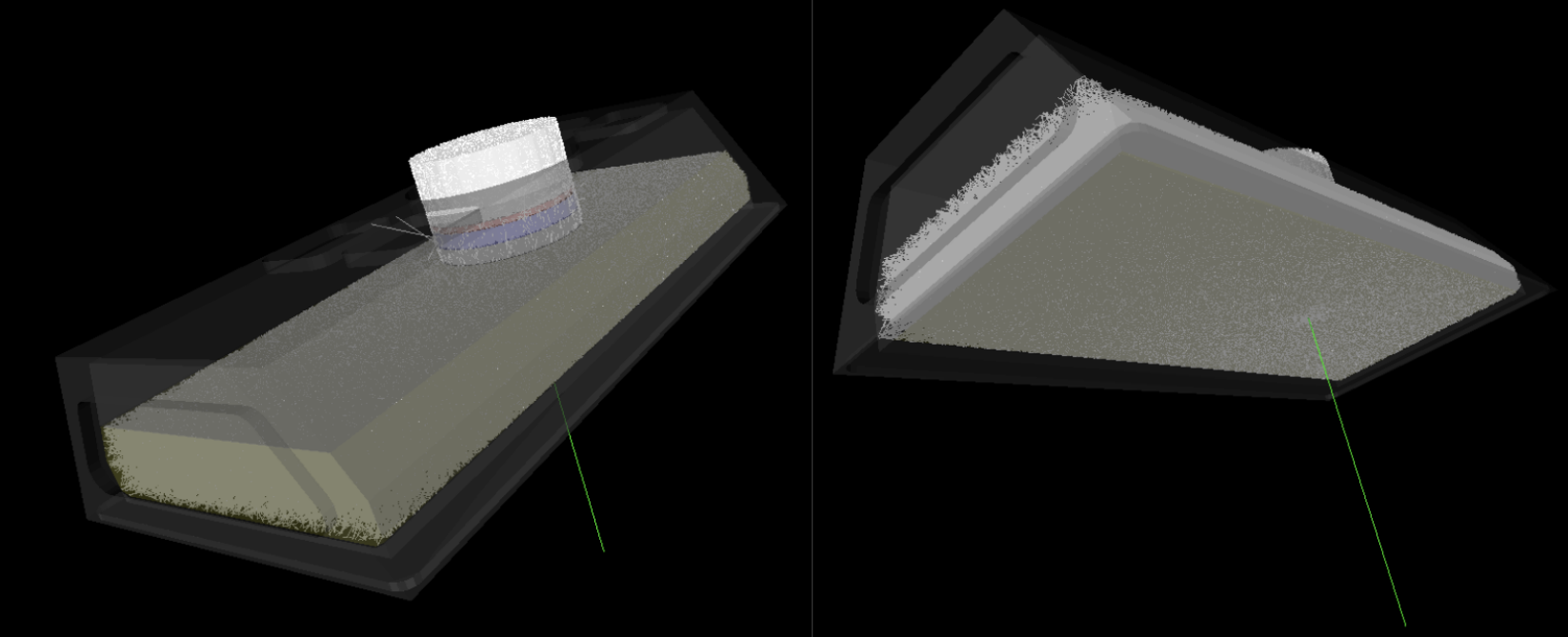}
   \end{tabular}
   \end{center}
   \caption 
   { \label{fig:claire_sim} 
Snapshots of the Geant4 simulation: the green line is the gamma photon which hits the aluminum case, enters the CsI and produces scintillation photons, whose trajectories are so crowded that they almost fill the crystal.}
   \end{figure}

We simulated 1000 X-ray photons of 122 keV hitting the CsI block in each position as shown in Fig. \ref{fig:claire_sim}. Then, for each run, we considered the number of optical photons which were absorbed in the photocathode and normalized it to the sum of all the 40 measurements. We will refer to such quantity as PMT relative response. Although we are not including the PMT electronic readout (the PMT is simulated just as an empty volume in BoGEMMS-HPC), we assume that the actual response of the PMT (which would be expressed in Volt) is related to the number of absorbed optical photons in the photocathode through a proportionality factor, which becomes irrelevant when dealing with the relative response (i.e. normalizing to the sum of all the measurements). \\
The number of photons absorbed by the photocathode in the simulation is assumed to follow a Poisson distribution, i.e. if we count $N$ photons being absorbed in the photocathode, the associated error is $\sqrt{N}$. 
In order to best reproduce the experiment, only events in which the X-ray photon entirely deposits its energy in scintillation photons had to be taken into account. To do so, for each beam position we collected the total energy deposit in the photocathode for each input X-ray. Only the events whose total energy deposit lies within $\pm 2$ standard deviations from the mean were kept, all the others being associated to partial conversions of the primary X-ray photon. Having 1000 input events ensured enough statistics to perform this kind of analysis. \\
As discussed in Sec. \ref{sec:opt_phys}, Geant4 allows the setting of various parameters to specify the optical properties of the surfaces, e.g. the roughness, and how optical photons interact with them. These parameters highly influence the simulation outcome and its comparison with experimental data, hence their proper definition and understanding become crucial for our purposes. We then performed several comparisons for different combinations of those parameters, in particular those corresponding to the CsI - Teflon boundary, and tried to select the set-up which best reproduces the data; in addition, the effect of changing the absorption length of the CsI crystal was also studied.

\subsection{Results}
\label{sec:res_claire}
\begin{figure}
   \begin{center}
   \begin{tabular}{c} 
   \includegraphics[height=6cm]{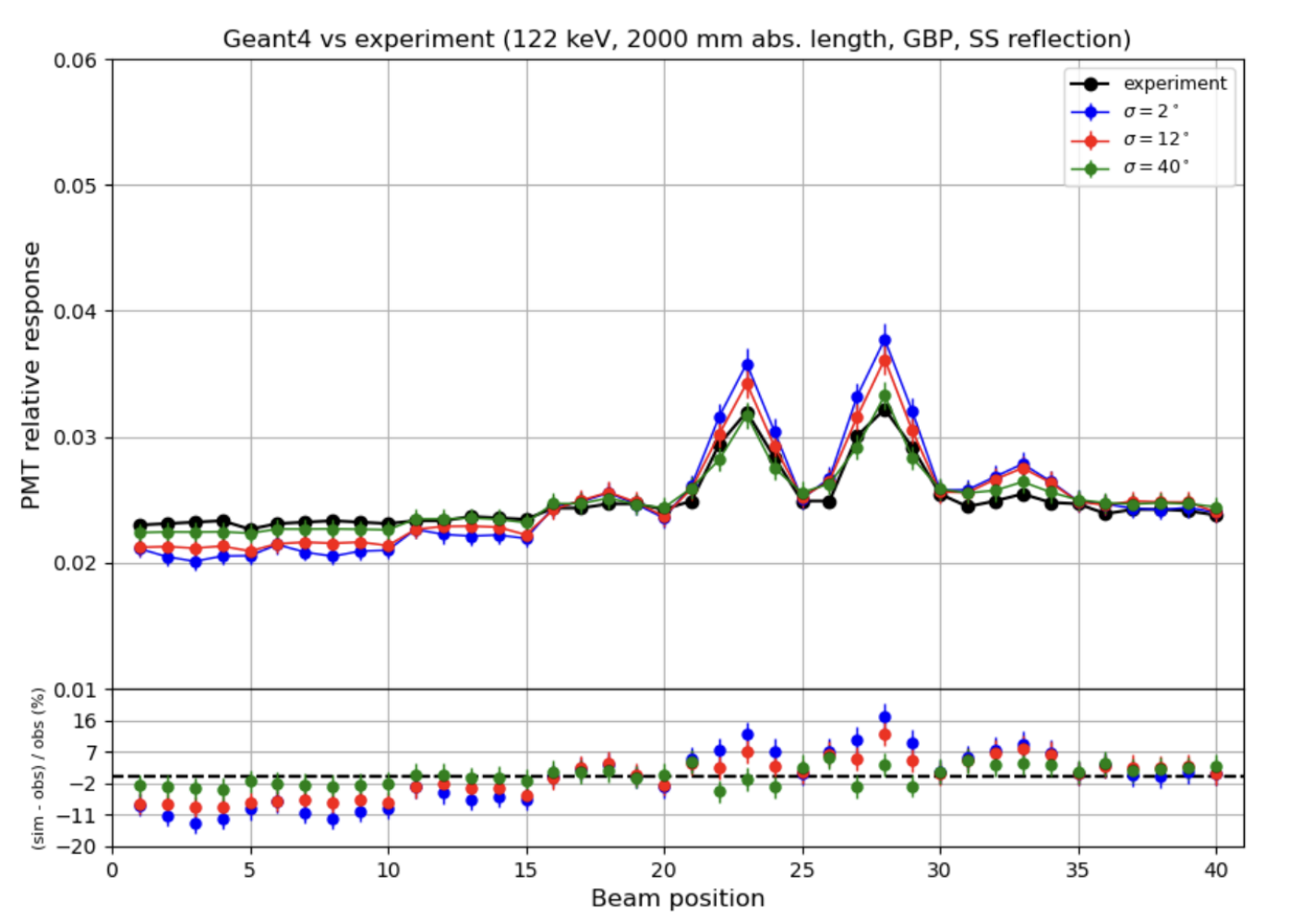}\includegraphics[height=6cm]{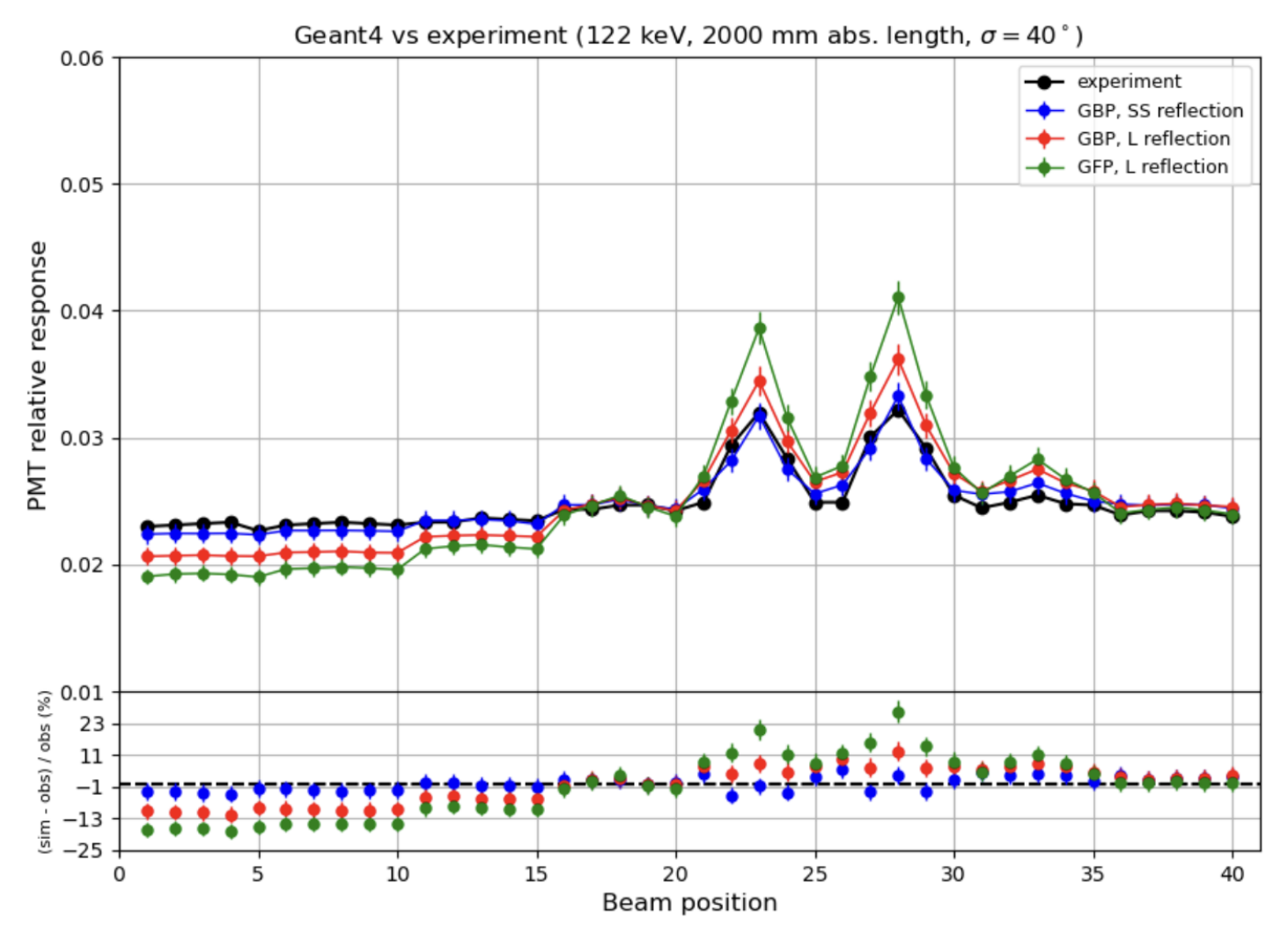}
   \end{tabular}
   \end{center}
   \caption 
   { \label{fig:comp_sigma_surf} 
Left panel: PMT relative responses from the simulation for each beam position, for a GBP surface, 100\% SS reflection and different values of $\sigma_\alpha$ (see the legend). Right panel: PMT relative responses from the simulation for each beam position, for $\sigma_\alpha$ = 40° with a different type of surfaces and reflection (see the legend). For both panels, the black points refer to experimental data and the bottom panels show the residuals for the simulations with respect to experimental data.}
   \end{figure}

\begin{figure}
   \begin{center}
   \begin{tabular}{c} 
   \includegraphics[height=6cm]{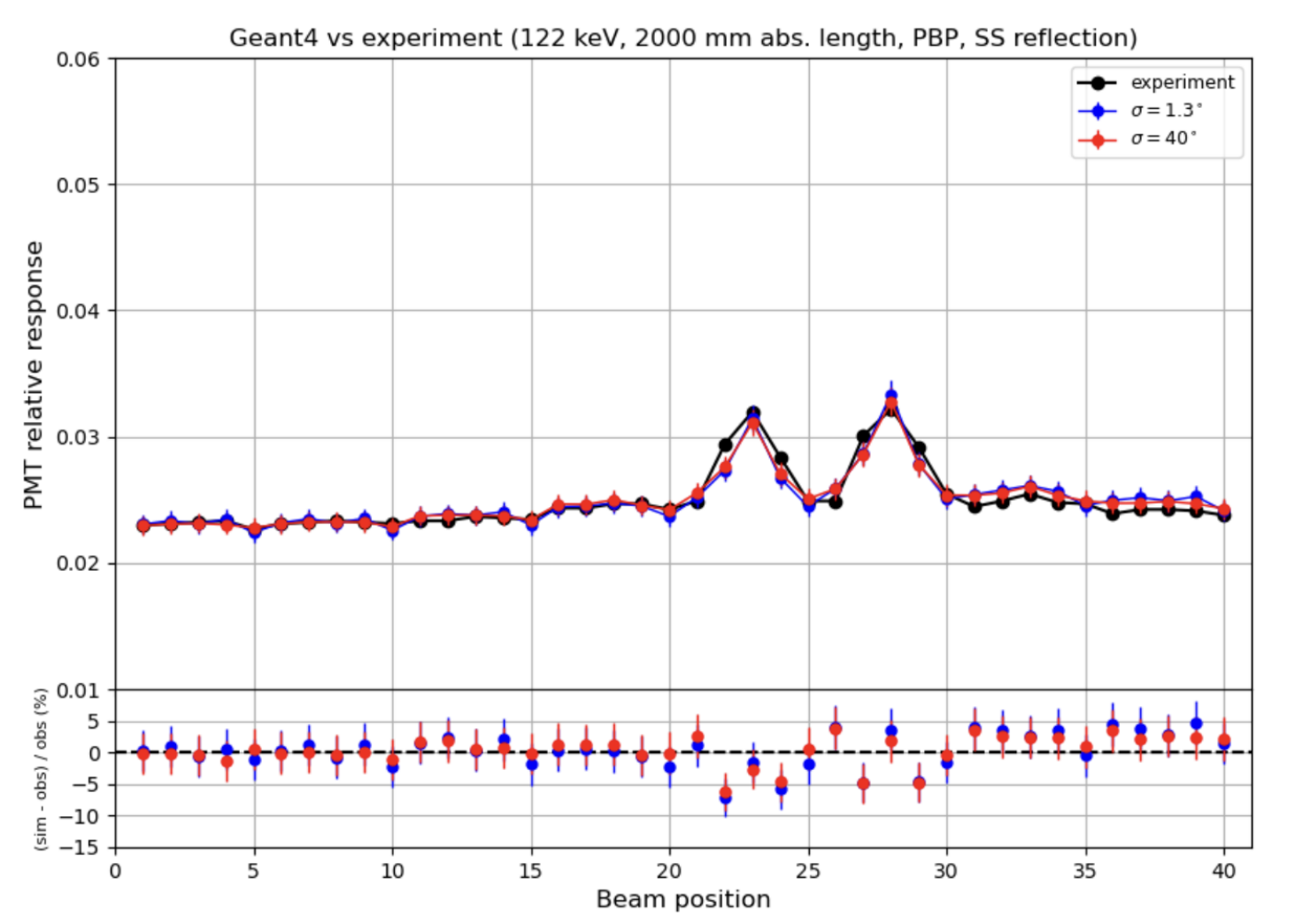}\includegraphics[height=6cm]{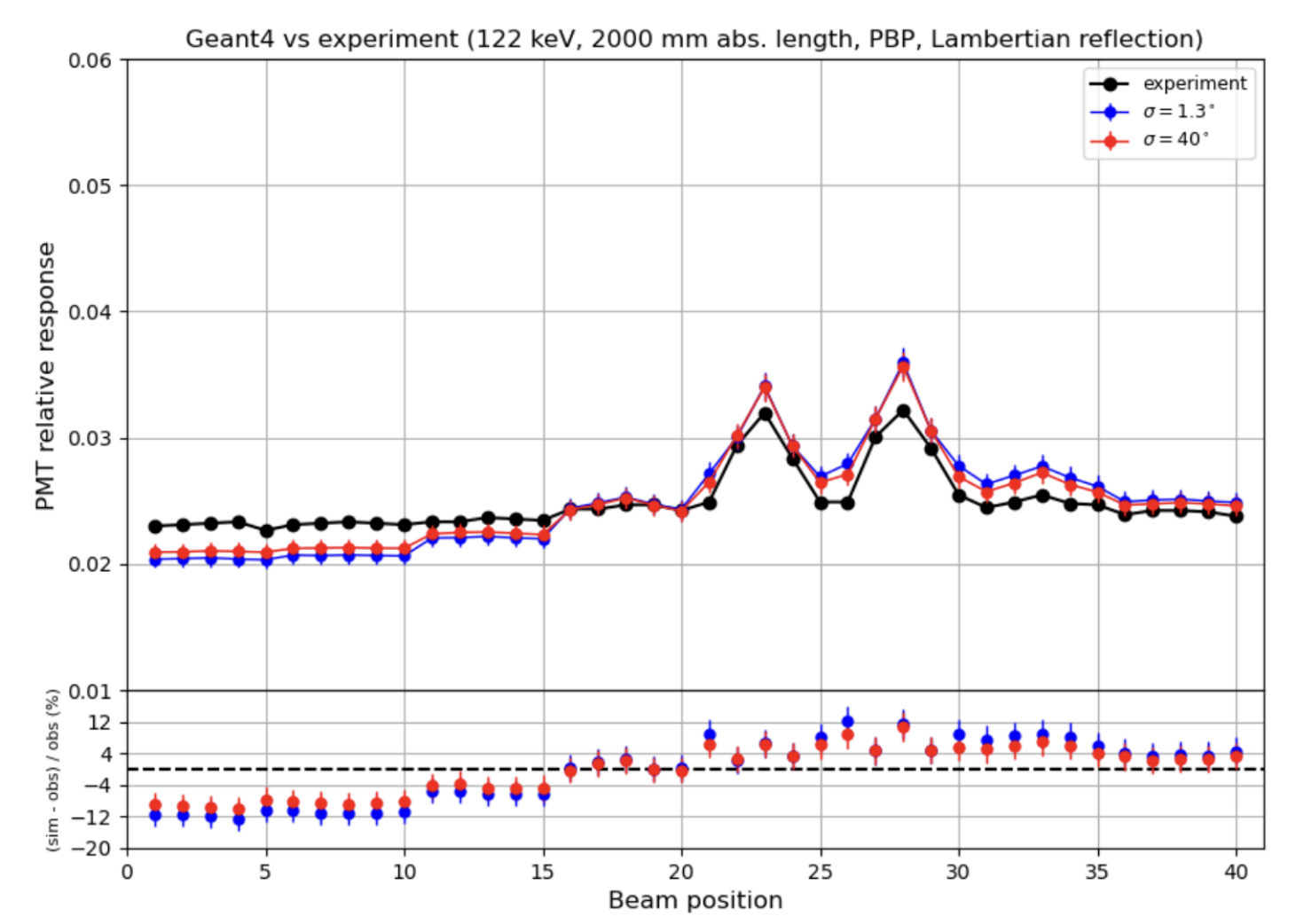}
   \end{tabular}
   \end{center}
   \caption 
   { \label{fig:comp_SS_L} 
Left panel: PMT relative responses from the simulation for each beam position, for a PBP surface, 100\% SS reflection and different values of $\sigma_\alpha$ (see the legend). Right panel: PMT relative responses from the simulation for each beam position, for a PBP surface, 100\% L reflection and different values of $\sigma_\alpha$ (see the legend). For both panels, the black points refer to experimental data and the bottom panels show the residuals for the simulations with respect to experimental data.}
   \end{figure}

\begin{figure}
   \begin{center}
   \begin{tabular}{c} 
   \includegraphics[height=6cm]{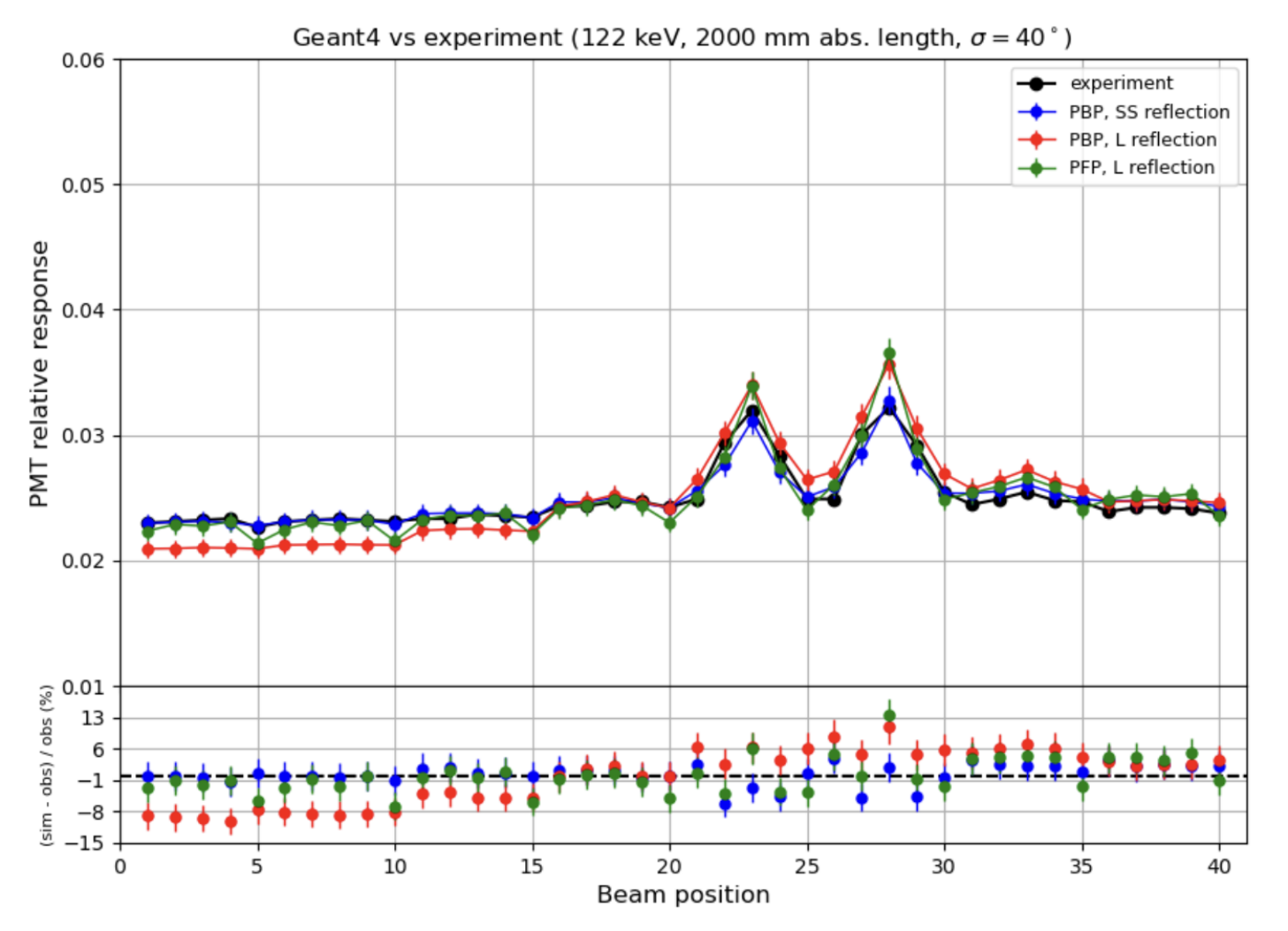}\includegraphics[height=6cm]{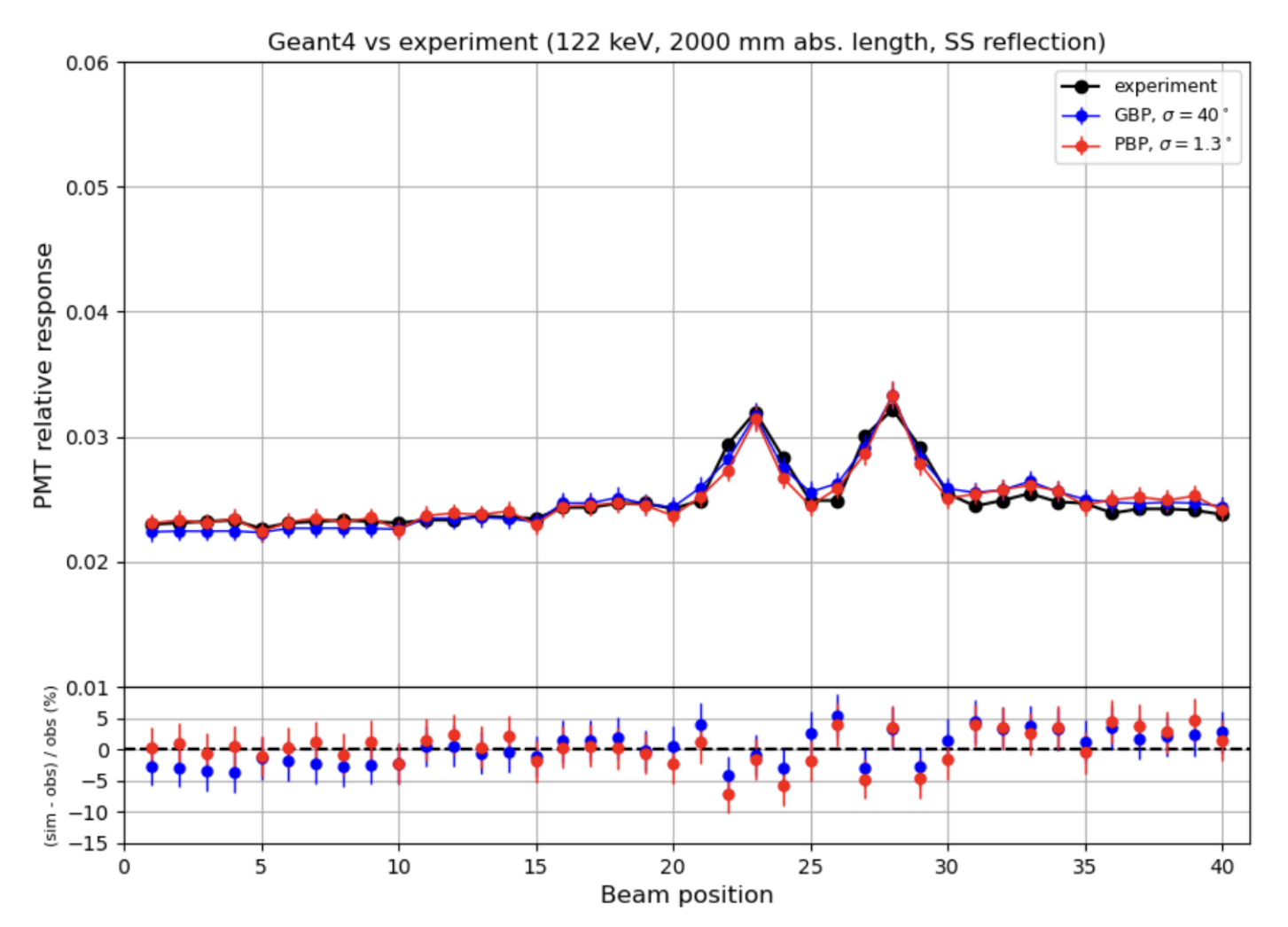}
   \end{tabular}
   \end{center}
   \caption 
   { \label{fig:comp_refl_ground_polish} 
Left panel: PMT relative responses from the simulation for each beam position, for $\sigma_\alpha$ = 40° with different type of surfaces and reflection (see the legend). Right panel: PMT relative responses from the simulation for each beam position, for 100\% SS reflection, $\sigma_\alpha$ = 40° and for GBP, PBP surfaces (see the legend). For both panels, the black points refer to experimental data and the bottom panels show the residuals for the simulations with respect to experimental data.}
   \end{figure}

\begin{figure}
   \begin{center}
   \begin{tabular}{c} 
   \includegraphics[height=6cm]{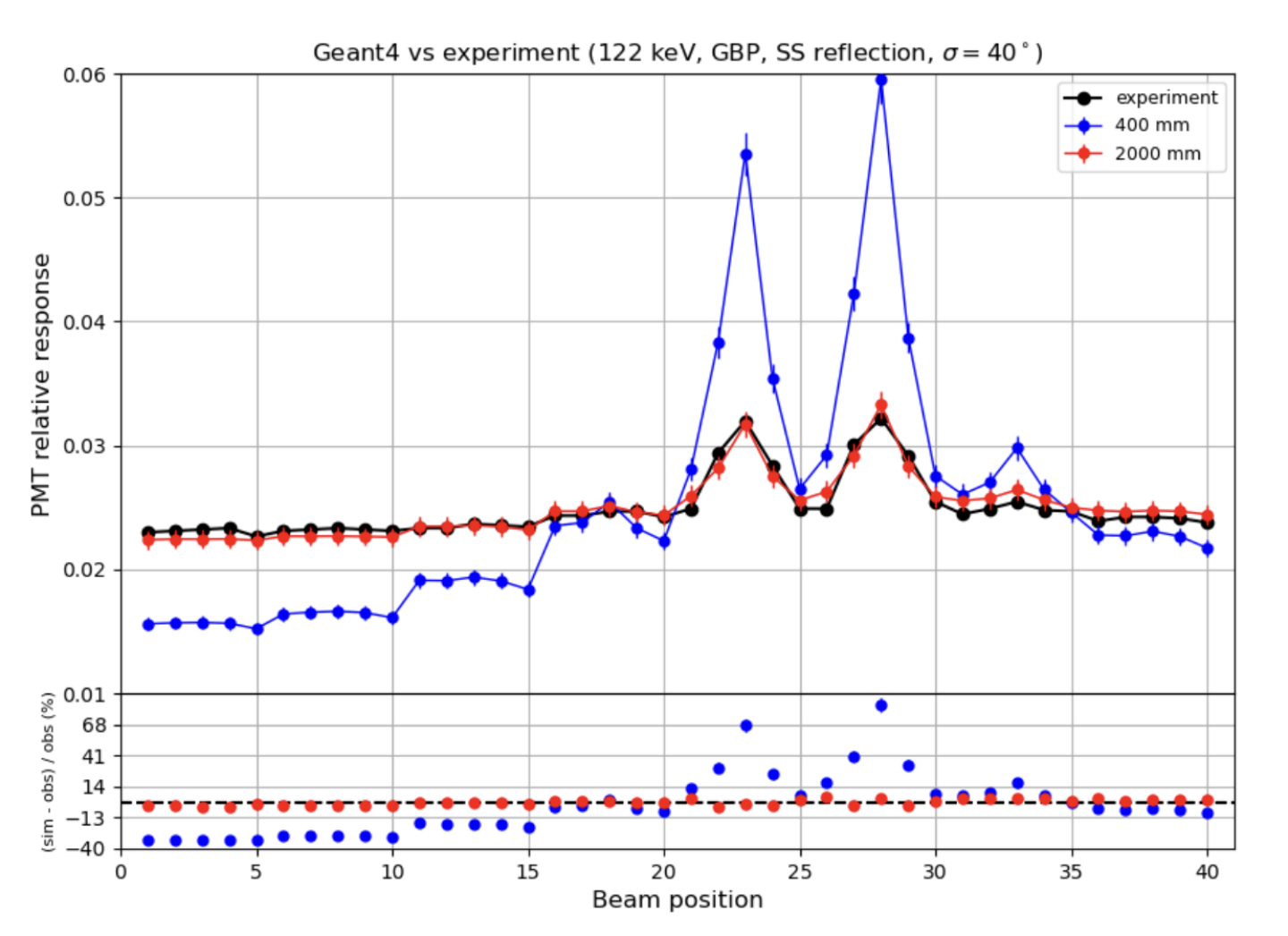}\includegraphics[height=6cm]{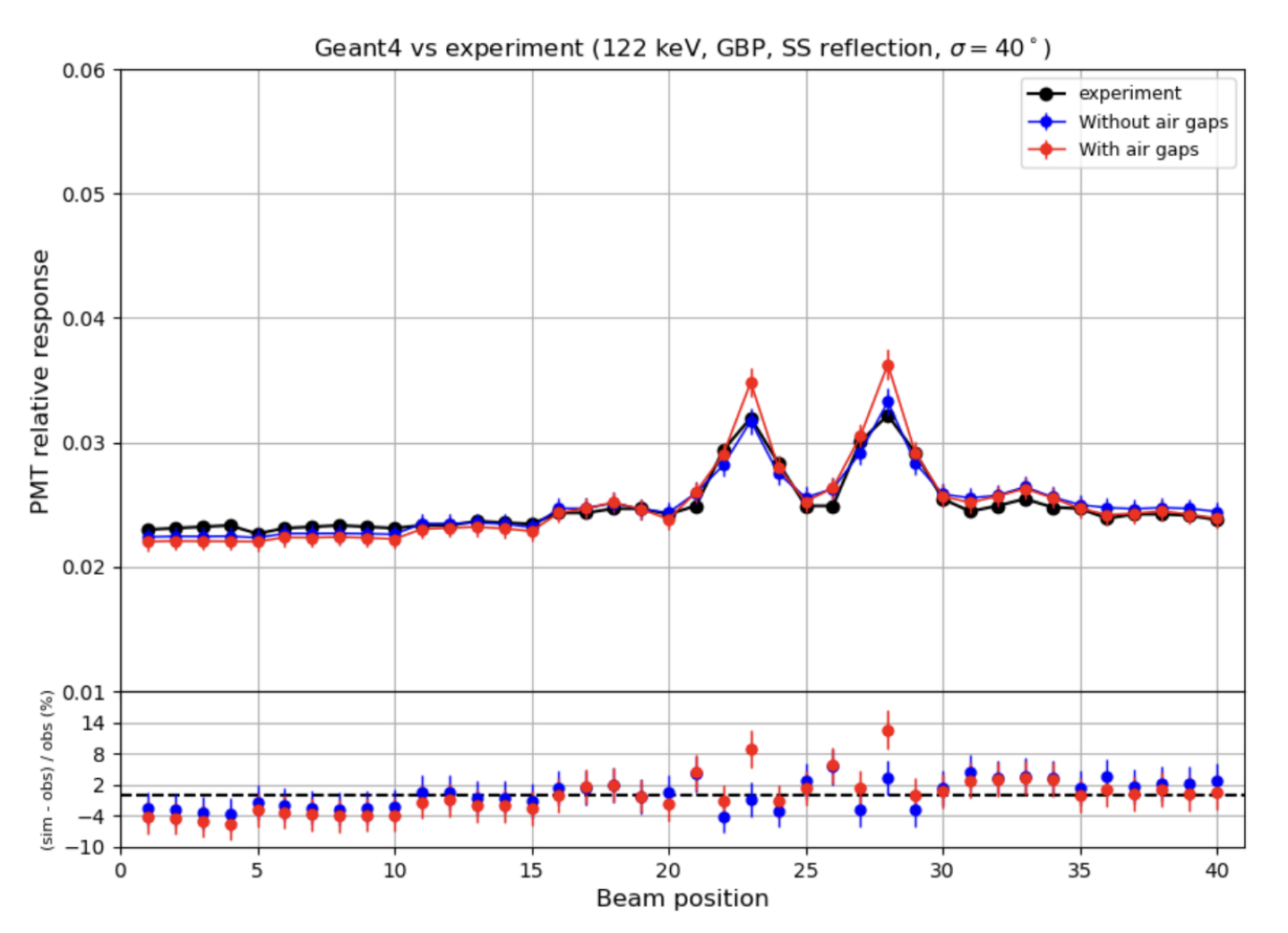}
   \end{tabular}
   \end{center}
   \caption 
   { \label{fig:comp_abslen_air} 
Left panel: PMT relative responses from the simulation for each beam position, for a GBP surface, 100\% SS reflection, $\sigma_\alpha$ = 40° and for absorption lengths of 400 mm, 2000 mm (see the legend). Right panel: PMT relative responses from the simulation for each beam position, for a GBP surface, 100\% SS reflection, $\sigma_\alpha$ = 40° with and without air gaps above and below the SiPad (see the legend). For both panels, the black points refer to experimental data and the bottom panels show the residuals for the simulations with respect to experimental data.}
   \end{figure}

We present here the results of the simulations for the relative response of the PMT for each position, along with the experimental data for comparison. We perform the comparison for several combinations of the optical parameters to find the set-up that best matches the data. For each of the following cases, we set the parameter Reflectivity to 0.99\cite{Janecek:2012} and the absorption length to 2 m for the CsI crystal\cite{Alvarez:2022}. \\
We first compare the results for ground painted surfaces between CsI and Teflon; at first, we study the effect of having different levels of roughness, which is parametrised by $\sigma_\alpha$. The left panel of Fig. \ref{fig:comp_sigma_surf} shows the relative responses for each position with a GroundBackPainted (GBP) surface, 100\% specular spike (SS) reflection and three different values of $\sigma_\alpha$ (2$^\circ$, 12$^\circ$, 40$^\circ$), together with the experimental data. It is quite clear that having a higher roughness corresponds to a better agreement with the data. In particular, better agreement is achieved for $\sigma_\alpha$ = 40$^\circ$, where the deviation from the experiment is, on average, below $\sim$5\%. For $\sigma_\alpha$ = 12$^\circ$ the deviations increase up to $\sim10$\%, while for $\sigma_\alpha$ = 2$^\circ$ they lie near 10\% on average with peaks of $\sim15$\% for beams 23 and 28 (i.e. the ones closer to the PMT). The Lambertian (L) reflection probability can be set for GroundBackPainted surfaces, whereas for GroundFrontPainted (GFP) only Lambertian reflection can occur. The right panel of Fig. \ref{fig:comp_sigma_surf} shows the relative responses for a fixed $\sigma_\alpha$ = 40$^\circ$, for both GBP and GFP, with 100\% Lambertian reflection; for comparison, also the previous case with 100\% SS reflection with $\sigma_\alpha$ = 40$^\circ$ is shown, along with the experimental data. The result with SS reflection represents the best agreement with the data; changing reflection to the Lambertian type slightly worsens the comparison for GBP, with larger deviations of a few per cent, while selecting a GFP surface results in a larger discrepancy ($\sim15$\% deviation far from the PMT and $\sim25$\% nearby).
In the left panel of Fig. \ref{fig:comp_SS_L}, we show the results for a PolishedBackPainted (PBP) surface for two different levels of roughness, a very low one ($\sigma_\alpha$ = $1.3^\circ$ as applied in \citenum{Janecek:2010}) to best mimic the polished nature of the surface, and a higher one ($\sigma_\alpha$ = 40$^\circ$). Note that the roughness here is actually associated with the CsI-gap interface and not with the wrapping, which is instead perfectly smooth. Increasing the roughness does not significantly impact the distribution, contrary to the ground surfaces, and both cases show a very good agreement with the data (deviations below $\sim2$\% far from the PMT and $\sim5$\% nearby). The right panel of Fig. \ref{fig:comp_SS_L} shows the case with PBP and 100\% Lambertian reflection. As for the PBP with SS reflection, different levels of roughness provide similar results within the errors. However, a higher roughness results in a better agreement with the experiment for beam positions far from the PMT, with respect to the case with smaller $\sigma_\alpha$. Still, the Lambertian reflection generally worsens the comparison with the data with respect to having 100\% SS reflection, also for polished painted surfaces, since the deviations from the experiment are roughly 10\% either far from or near the PMT.
In left panel of Fig. \ref{fig:comp_refl_ground_polish}, we compare the relative responses fixing $\sigma_\alpha = 40^\circ$, for PBP surfaces with in one case SS reflection and the other with L reflection, including also the case with PFP surface with Lambertian reflection (which in this case is the only option). As for ground surfaces, SS reflection provides the best results, with discrepancies which are lower than a few per cent than those of the other cases. It is also interesting to compare ground and polished painted surfaces, in particular the cases with the best agreement with the data for both the categories: (polished and ground) back painted surfaces, with SS reflection and with $\sigma_\alpha = 1.3^\circ$ and $\sigma_\alpha = 40^\circ$ respectively. As shown in the right panel of Fig. \ref{fig:comp_refl_ground_polish}, the two results are very similar and both are well in agreement, within the statistical error, with the experimental data.
So far, for all the cases we kept the absorption length of the CsI crystal fixed to the value of 2 m. In the following, we test a different choice. In the left panel of Fig. \ref{fig:comp_abslen_air}, we compare the results for GBP surface, SS reflection and $\sigma_\alpha = 40^\circ$ with two different absorption lengths: 2 m (the one used so far) and 40 cm. We can immediately conclude that a shorter absorption length dramatically worsens the agreement with the experiment (the discrepancy is almost 40\% far from the PMT and even $\sim70$\% for positions 23 and 28), thus ensuring that selecting an absorption length of 2 m represents a good choice. Finally, we investigate the possibility of imperfect coupling between the CsI and PMT by introducing two air gaps above and below the SiPad (both with a thickness of 0.01 mm). We show here, in the right panel of Fig. \ref{fig:comp_abslen_air}, the results for GBP, SS reflection and $\sigma_\alpha = 40^\circ$ in one case with air gaps, and in another case without them. Having the air gaps results in a worse agreement with the experimental data since the PMT relative response is overestimated for positions in front of the PMT (23 and 28) and slightly suppressed for far positions. A possible explanation is that the optical photons that originate from positions far from the PMT have more chance to enter the SiPad with a larger incidence angle with respect to positions near the PMT, hence being reflected back from one of the two air gaps with higher probability. \\
Among all the parameter combinations we tested, two of them gave the most satisfactory results: the GroundBackPainted and the PolishedBackPainted surfaces, both with 100\% specular spike reflection and with $\sigma_\alpha$ = 40$^\circ$ and $\sigma_\alpha$ = 1.3$^\circ$, respectively. However, the case with $\sigma_\alpha$ = 1.3$^\circ$ represents a more realistic scenario than $\sigma_\alpha$ = 40$^\circ$. A surface with  $\sigma_\alpha$ = 40$^\circ$ is extremely rough, and the suspect is that the GBP case matches the experiment only if one artificially tunes the roughness to such high levels.

\section{TOWARDS THE RESPONSE FUNCTION FOR THE COSI ACS}
\label{sec:towards}
The verification of the Geant4 optical physics library represents a first step towards having validated Geant4 simulations for the construction of a response function of the COSI ACS. The mechanism through which optical photons propagate and interact before reaching the readout device has an important contribution for the overall detection efficiency and resolution of the ACS. The interaction position of the high-energy photons (or particles) with the ACS affects the probability for the generated optical photons to reach the readout device. This may cause a non-uniform light collection efficiency and energy threshold across the ACS surface. Also, the statistical fluctuation in the generation and detection of optical photons introduces an additional contribution in the ACS energy resolution, that can in turn have a position dependence too. We then
propose the use of a response function that encodes the light collection efficiency and energy resolution in a 2D binned map. Anticipating calibration measurements for the new COSI SMEX Engineering Model on which to validate simulations, we present a preliminary characterization of the response and energy resolution for the updated ACS design (BGO + SiPMs) using the optical parameters derived from the CLAIRE experiment (see Sec. \ref{sec:claire}), i.e. BGO polish surface, specular reflection and absorption length of 200 cm. We compute the light collection (i.e. the number of detected optical photons) in 50 squared bins of the ACS surface. The energy resolution is instead evaluated for the full panel, although we plan to characterize its spatial distribution. 

\subsection{COSI SMEX ACS light collection map and energy resolution}
\begin{figure}
   \begin{center}
   \begin{tabular}{c} 
   \includegraphics[height=5cm]{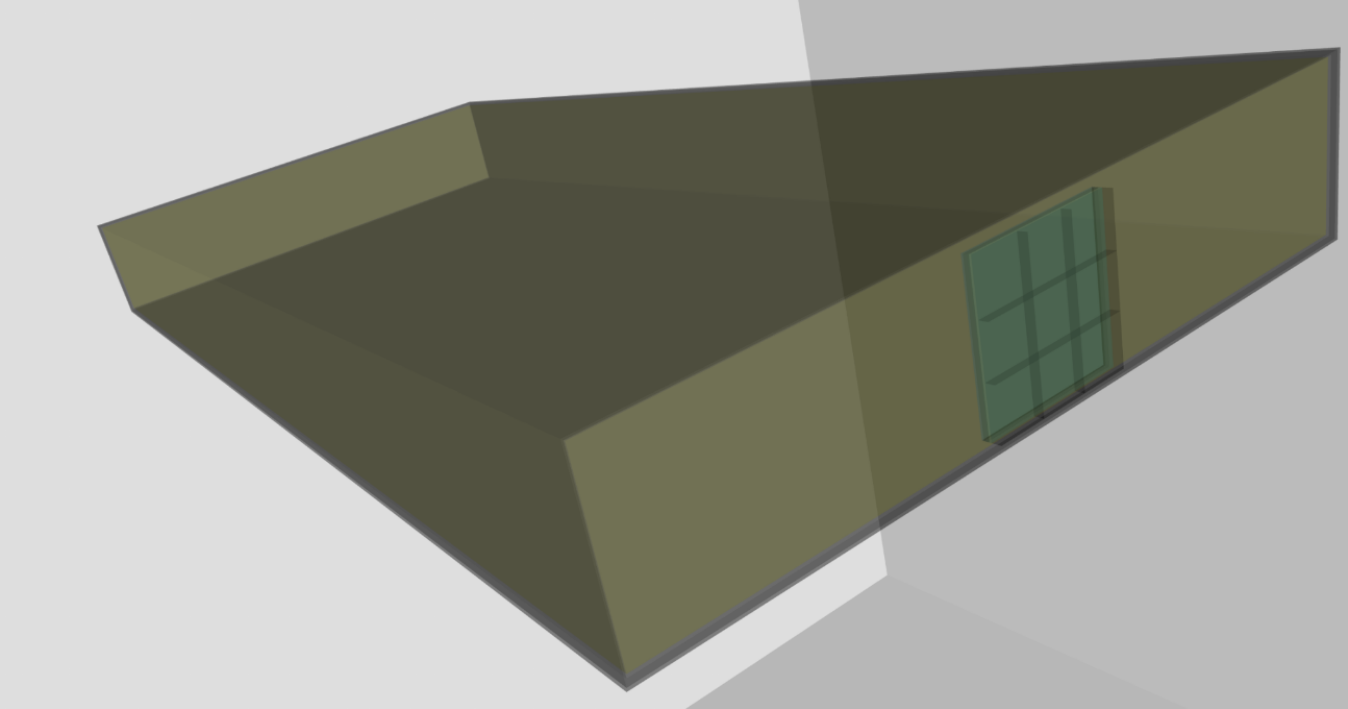}
   \end{tabular}
   \end{center}
   \caption 
   { \label{fig:SiPM} 
Geant4 mass model of the COSI SMEX ACS prototype with BGO and SiPMs. The SiPMs are visible as a green 3x3 grid placed on the side.}
   \end{figure}

\begin{figure}
   \begin{center}
   \begin{tabular}{c} 
   \includegraphics[height=6cm]{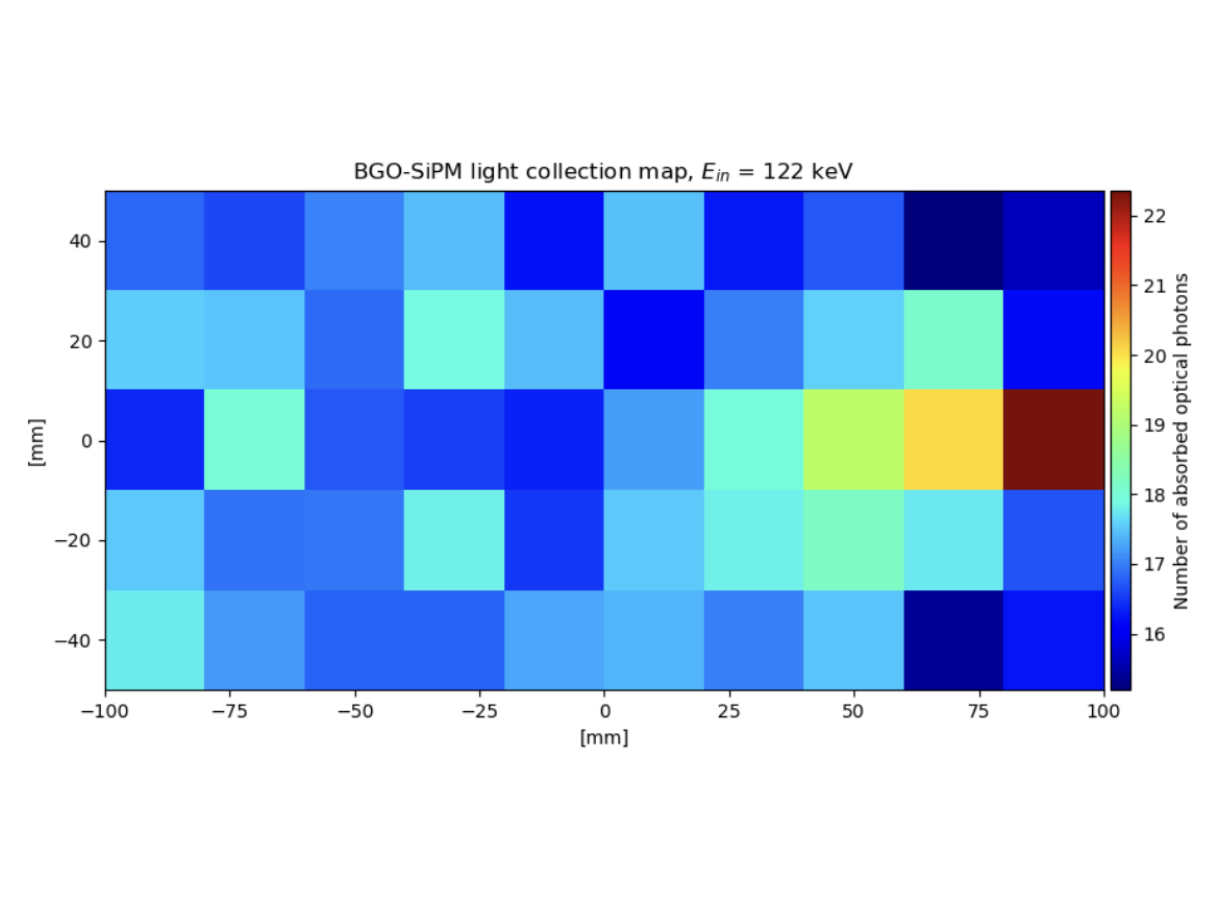}\includegraphics[height=6cm]{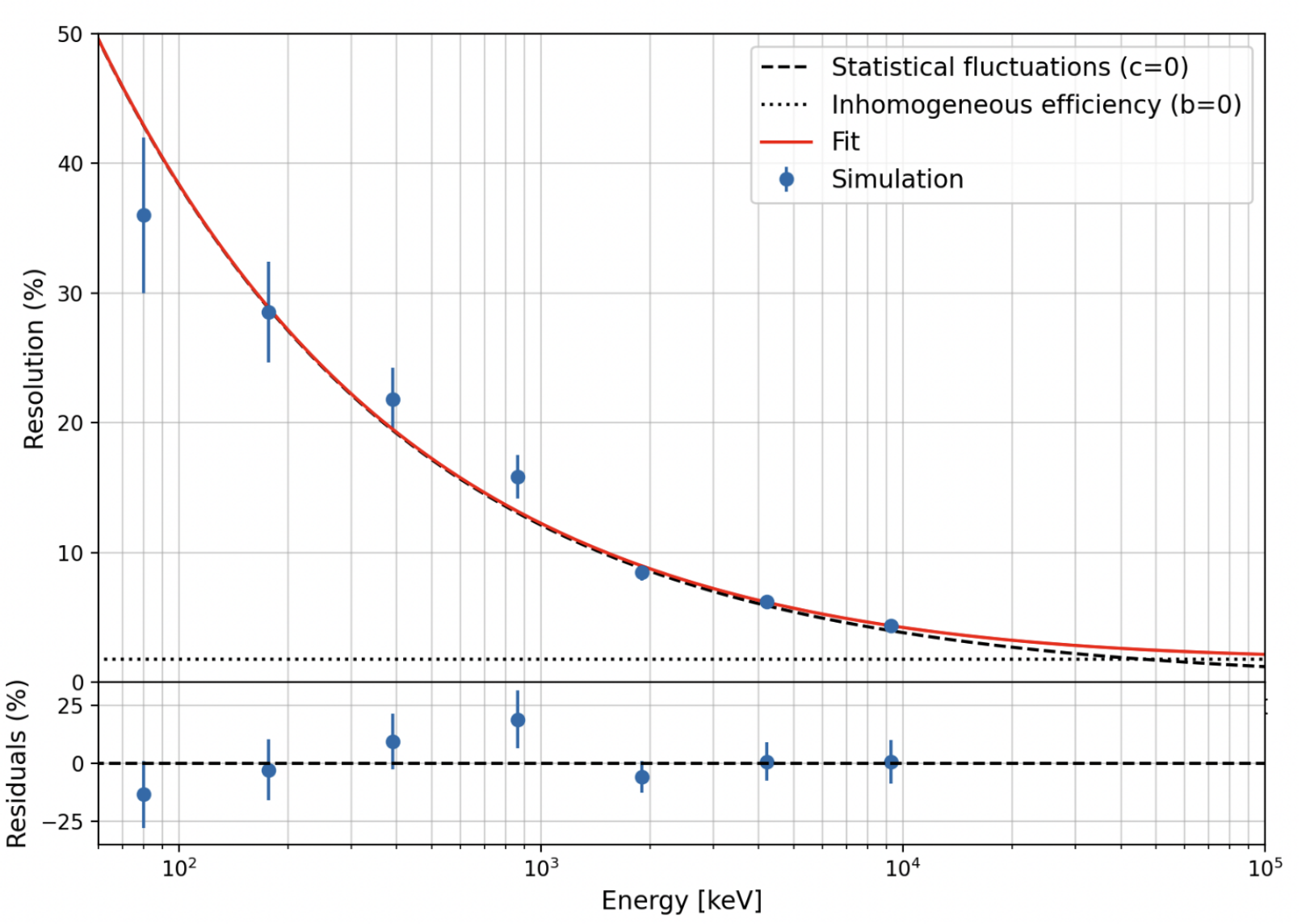}
   \end{tabular}
   \end{center}
   \caption 
   { \label{fig:response_resolution} 
Left panel: light collection map for the  ACS panel, with the SiPM array being on the right side. Right panel: energy resolution for a set of energies, with the fit model ($b = 383.71 \pm 20.17$, $c = 1.79 \pm 1.11$) plotted in red.}
   \end{figure}

We build a first COSI SMEX ACS prototype (Fig. \ref{fig:SiPM}) consisting of a 20x10x2 cm$^3$ block of BGO, wrapped in reflective tape and coupled with a 3x3 array of SiPMs centered on the 10x2 cm$^2$ face. Each SiPM unit (onsemi J series 60035 model) is a 6x6 mm$^2$ square. A pad of EJ-560 optically couples the BGO with the SiPM array. The SiPMs are simulated as empty, fully absorbing volumes. The actual number of detected photons is obtained by applying to the number of absorbed photons the SiPM photon detection efficiency (PDE), that accounts for the quantum efficiency, the avalanche initiation probability and the geometric filling factor. The PDE is a function of the optical photon energy and has a mean value of $\sim20\%$. Note that with the SiPMs we do not have problems of thin layer as in the BGO-PMT interface, as the actual chance of detecting an optical photon is directly encoded in the PDE.
The full panel is illuminated by a parallel beam of monochromatic X-ray photons, and the number of detected optical photons is recorded. We divided the BGO surface into 50 bins (10x5) and evaluated the average number of optical photons detected by the SiPMs for each bin. The light collection map for 122 keV input photons is shown in the left panel of Fig. \ref{fig:response_resolution}, for a BGO absorption length of 200 cm and with SiPMs placed on the right side. The light collection is overall homogeneous across the panel surface, with the exception of the bin right in front of the SiPMs where the response is enhanced. The energy resolution is computed for a few energies ranging from 80 keV to 10 MeV and shown in the right panel of Fig. \ref{fig:response_resolution}. It is defined as the full width at half maximum (FWHM) normalized to the peak energy. For the fit we follow the same parametrization for the resolution used for the Fermi GBM detector\cite{Bissaldi:2008df}:
\begin{equation}
    \text{FWHM} = \sqrt{a^2 + b^2\,E + c^2\,E^2}\,\,,
    \label{eq:fwhm}
\end{equation}
where $a$ accounts for the limiting electronic resolution, $b$ for the statistical fluctuation of the detected photons, and $c$ for the inhomogeneous response of the ACS panel. We set $a=0$, as the simulation does not account for the electronic resolution. 
We obtain a resolution of $\sim15$\% @ 662 keV, similar to the values reported for the GBM instrument. In the right panel of Fig. \ref{fig:response_resolution} we also show the two singular components contributing to the resolution, i.e. the statistical fluctuations (setting $c=0$) and inhomogeneous efficiency (setting $b=0$). The statistical factor is the most important contribution and it is responsible for the increase at low energies. The inhomogeneous efficiency behaves as constant contribution and starts to become relevant at higher energy ($E \gtrsim 40$ MeV), acting as a lower limit for the total resolution.

\section{CONCLUSION AND FUTURE WORK}
\label{sec:conclusions}
We presented a verification of the Geant4 optical physics simulation against analytical models for the transmission, reflection and absorption and laboratory measurements. We showed that Geant4 is capable of correctly reproducing the theoretical expected number of transmitted, reflected and absorbed optical photons out of a 5-layer system, emulating the BGO-PMT interface (see Sec. \ref{sec:an_ver}), with an overall accuracy of $<15\%$. In particular, the number of absorbed photons, relevant for predicting the number of photoelectrons emitted in the PMT, is simulated with a discrepancy of $\sim8\%$ from the analytical model. The Geant4 optical physics was further tested by comparison with experimental data, involving the response measurement of the ACS prototype with CsI and PMTs (see Sec. \ref{sec:claire}). This work also allowed us to investigate the impact of the main optical parameters and to select those producing the best match with the experiment. Choosing an absorption length of 200 cm and a polished surface for the CsI, and 100\% specular reflection from the reflective layer, Geant4 reproduced the experimental response with an overall accuracy of $<5\%$, for all hitting positions of 122 keV photons. Finally, we presented a preliminary characterization of the light collection map and energy resolution for an updated ACS prototype with BGO and SiPMs (see Sec. \ref{sec:towards}), using the same optical setup found for the CLAIRE experiment. The light collection obtained with 122 keV input photons featured a global homogeneity across the BGO surface, with an enhanced response right close to the SiPMs. We computed the energy resolution for a few energies ranging from 80 keV to 10 MeV and found a resolution of $\sim15\%$ @ 662 keV. The obtained trend is similar to what found for the Fermi GBM BGO detector. \\ Calibration measurements for the COSI SMEX ACS are currently ongoing to benchmark the Geant4 simulations and drive the evaluation of the response function. The results will be implemented in the detector effects engine iteratively improving the accuracy of the ACS simulations.

\acknowledgments

This work is supported by Centro Nazionale di Ricerca in High-Performance Computing, Big Data and Quantum Computing (CN\_00000013 - CUP C53C22000350006). This work is also supported in part by CNES.

% References
\bibliography{bibtex} % bibliography data in report.bib
\bibliographystyle{spiebib} % makes bibtex use spiebib.bst

\end{document}